\documentclass{aa}  
\usepackage{graphicx}
\usepackage{txfonts}
\usepackage{natbib}
\usepackage{hyperref}
\bibpunct{(}{)}{;}{a}{}{,} 

\usepackage{graphicx}
\usepackage{xcolor}
\newcommand{\Ni}{{$^{56}$Ni}}
\newcommand{\ms}{$M_{\odot}$}
\newcommand{\rs}{$R_{\odot}$}
\newcommand{\mzams}{$M_{\rm ZAMS}$}
\newcommand{\mni}{$M_{\rm Ni}$}
\newcommand{\mix}{$^{56}\rm Ni$ mixing}
\newcommand{\te}{$t_{\rm exp}$}

\begin{document}

	\title{Progenitor properties of type II supernovae: fitting to hydrodynamical models using Markov chain Monte Carlo methods}
	
	\author{L. Martinez \inst{1,2,3}
			\and M.~C. Bersten \inst{1,2,4}
			\and J.~P. Anderson \inst{5}
			\and S. González-Gaitán \inst{6}
			\and F. Förster \inst{7,8,9}
			\and G. Folatelli \inst{1,2,4}}

	\institute{Instituto de Astrofísica de La Plata (IALP), CCT-CONICET-UNLP. 
			Paseo del Bosque S/N, B1900FWA, La Plata, Argentina\\
			\email{laureano@carina.fcaglp.unlp.edu.ar}
		\and
			Facultad de Ciencias Astronómicas y Geofísicas,
			Universidad Nacional de La Plata, Paseo del Bosque S/N, B1900FWA, La Plata, Argentina
		\and
		    Universidad Nacional de Río Negro. Sede Andina, Mitre 630, (8400) Bariloche, Argentina
		\and
			Kavli Institute for the Physics and Mathematics of the Universe (WPI),
			The University of Tokyo, 5-1-5 Kashiwanoha, Kashiwa, Chiba 277-8583, Japan
		\and
			European Southern Observatory, Alonso de Córdova 3107, Casilla 19, Santiago, Chile
		\and
			CENTRA-Centro de Astrofísica e Gravitaçäo and Departamento de Física, Instituto Superior 					Técnico, Universidade de Lisboa, Avenida Rovisco Pais, 1049-001 Lisboa, Portugal
		\and
			Centre for Mathematical Modelling, University of Chile, Santiago, Chile
		\and
			Millennium Institute of Astrophysics, Casilla 36-D, 7591245 Santiago, Chile
		\and
			Departamento de Astronomía, Universidad de Chile, Camino El Observatorio 1515, Las Condes, 					7591245 Santiago, Chile\\}
%

\titlerunning{Progenitor properties of SNe~II}


\date{Received XXX; accepted XXX}
 


\abstract 
{The progenitor and explosion properties of type II supernovae (SNe~II) are fundamental to understand the evolution of massive stars. Special interest has been given to the range of initial masses of their progenitors, but despite the efforts made, it is still uncertain. Direct imaging of progenitors in pre-explosion archival images point out an upper initial mass cutoff of $\sim$18~\ms. However, this is in tension with previous studies in which progenitor masses inferred by light curve modelling tend to favour high-mass solutions. Moreover, it has been argued that light curve modelling alone cannot provide a unique solution for the progenitor and explosion properties of SNe II.}
{We develop a robust method which helps us to constrain the physical parameters of SNe~II by fitting simultaneously their bolometric light curve and the evolution of the photospheric velocity to hydrodynamical models using statistical inference techniques.}
{Pre-supernova red supergiant models were created using the stellar evolution code MESA, varying the initial progenitor mass. The explosion of these progenitors was then processed through hydrodynamical simulations, where the explosion energy, synthesised nickel mass, and the latter's spatial distribution within the ejecta were changed. We compare to observations via Markov chain Monte Carlo methods.}
{We apply this method to a well-studied set of SNe with an observed progenitor in pre-explosion images and compare with results in the literature. Progenitor mass constraints are found to be consistent between our results and those derived by pre-SN imaging and the analysis of late-time spectral modelling.}
{We have developed a robust method to infer progenitor and explosion properties of SN II progenitors which is consistent with other methods in the literature, which suggests that hydrodynamical modelling is able to accurately constrain physical properties of SNe~II. This study is our starting point for a further analysis of a large sample of hydrogen-rich SNe.}


\maketitle

\section{Introduction} 
\label{sec:intro} 

It has been established that most of the stars with initial masses greater than 8\,$M_{\odot}$ finish their evolution in a violent explosion \citep{woosley+02,heger+03}, known as a core-collapse supernova (CCSN).
They are observationally classified according to their spectral characteristics \citep{filippenko97}. Type II supernovae (SNe II) show strong and prominent P-Cygni hydrogen lines. Subsequent division was introduced based on their light curve (LC) decline rates after maximum into II-Plateau (IIP) and II-Linear (IIL). However, recent studies have questioned this subdivision and propose the existence of a continuous sequence of LC slopes among SNe~II \citep{anderson+14,sanders+15,galbany+16,rubin+16}. Therefore, throughout this paper we use `SNe II' to refer to these two groups together.
A further classification exists among SNe that show hydrogen lines: the SN~1987A-like events, displaying unusually long-rising LCs \citep[e.g.][]{taddia+12,taddia+16}; the type IIn, showing narrow emission features in their spectra \citep[e.g.][]{schlegel90,taddia+13}; and the type IIb, which show hydrogen features at early times while later such lines disappear \citep[e.g.][]{filippenko+93}.
A recent study analysed the possible existence of a continuum between the SNe~II and IIb in terms of their photometric properties \citep{pessi+19}. However, they found clear differences between the two subgroups. These three groups show characteristics sufficiently distinct from SNe~II (as defined above) that we do not consider them in the present work. Only SNe~II are studied in this paper.

It is generally assumed that the progenitors of SNe~II are massive stars which have retained a significant fraction of their hydrogen-rich envelopes before explosion. In addition to this, it has been shown that these assumptions are consistent with detections of progenitor stars in pre-explosion images. These detections have constrained the progenitors to be red supergiant (RSG) stars in the zero-age main sequence (ZAMS) mass range of $\sim$8--20~\ms\ \citep{smartt15,davies+20}. 

Although SNe II are the most common type of SN in nature \citep{arcavi+10,li+11}, significant gaps remain in our knowledge of the different processes involved. Moreover, important discrepancies can be found in the literature regarding masses of the progenitors depending on the different methods used for the analysis.
Archival images provide the opportunity of detecting the progenitor star in images previous to the explosion. The luminosity and effective temperature of the progenitor can be obtained from fits to the spectral energy distribution or using bolometric corrections to convert single-band flux into luminosity \citep[see][among others]{vandyk+12a,davies+18}. Once the star is located in a Hertzsprung-Russell (HR) diagram, the mass of the star in the ZAMS is estimated by comparison with stellar evolutionary tracks. The acquisition of late-time imaging is the next step in the analysis of pre-explosion observations to confirm the progenitor identification through its disappearance \citep[][among others]{maund+14a,folatelli+16}. Despite being the most immediate method to determine which type of star produces each SN, this method can only be applied to the most nearby SNe (out to distances of $d \lesssim 30$~Mpc) because it requires images of high enough resolution and sensitivity.

Nebular-phase spectral modelling can also be used to constrain the progenitor masses of SNe. During the photospheric phase, the LC is mostly powered by reemission of the energy deposited by the shock wave. As time goes on, hydrogen recombination occurs at different layers of the object as a recombination wave moves inward. After the hydrogen has recombined, the envelope becomes transparent, the inner ejecta become visible and the nucleosynthesis yields can be analysed. Thus, using the dependency of oxygen production on progenitor initial mass, it is possible to distinguish between different progenitors \citep[see, e.g.][]{jerkstrand+12,jerkstrand+14}.

However, one of the most used methods to analyse the progenitor properties is the hydrodynamical modelling of SN LCs. It is well known that LCs are extremely sensitive to the physical properties of their progenitors (final masses and radii), as well as the properties of the explosion itself \cite[released energy, amount of synthesised radioactive nickel and its distribution, see for example][among others]{shigeyama+90,bersten+12}. The main problem of using this method is that sometimes LC modelling cannot provide a unique solution for the ejecta mass of SNe~II. There is a degeneracy among some progenitor properties when reproducing the observations. This means that progenitors with different physical properties can produce similar photometric and spectroscopic properties \citep{bersten+11,dessart+19,goldberg+19,goldberg+20}.

Following the degeneracies involved in constraining SN~II properties, the masses inferred by hydrodynamic simulations are usually much larger than those estimated from pre-SN imaging \citep[see][among others]{utrobin+09,utrobin+17}. 
Recently, \citet{morozova+18} and \citet{eldridge+19b} inferred initial masses for a group of SNe~II from LC modelling. They found that their results are mostly consistent with those from pre-explosion data. However, they did not take into account the ejecta velocities in their modelling and therefore the parameters derived are not unequivocally determined. Additionally, \citet[][MB19 hereafter]{martinez+19} presented hydrodynamic modelling of LCs and photospheric velocities of six objects with confirmation of the progenitor star and found that, in most cases, masses inferred by both methods were compatible. They also noted that the degeneracy in some physical parameters may be the reason for the differences found in the literature.
Hydrodynamical modelling can be applied to large distances and big samples as well, contrary to other methods which are more difficult to employ or present limitations due to distance. For this reason, efforts must be made in order to solve these issues and so be able to accurately constrain the physical and explosion properties of SNe~II.

In order to acquire a detailed outlook, we generate a grid of hydrodynamic models in the parameter space and a quantified fitting procedure. Similar grids and techniques were published in the recent works of \citet{morozova+18} and \citet{eldridge+19b}. The aim of the current work is to develop a procedure based on Markov chain Monte Carlo (MCMC) methods to construct the posterior distribution of the parameters involved given the LC and the photospheric velocity evolution together, and then constrain the progenitor properties of SNe~II.
In this paper we propose to check the robustness of this method with several tests. First, we derive the progenitor parameters of the same sample than in MB19 for comparison. 
In that work, the authors present detailed modelling of six objects using double polytropic progenitor structures and finding their optimal models by ``eye-fit'' to the observations. The differences in the current work are that we use pre-SN structures based on stellar evolution calculations and a more robust statistical analysis. The same hydrodynamic code is used in both cases.
The sample includes SN~2004A, SN~2004et, SN~2005cs, SN~2008bk, SN~2012aw, and SN~2012ec. We chose these objects as they represent some of the best-observed SNe II with enough photometric and spectroscopic monitoring during all the phases of evolution, detection of the progenitor star in pre-explosion images, and confirmation of the progenitor through its disappearance in late-time images.

We also test the validity of our method by comparing the initial masses derived from our fitting against those based on the analysis of the progenitor star in pre-explosion images for the same sample described above. 
For this purpose, we also include SN~2017eaw and SN~2018aoq as these are the last SNe II to be discovered and analysed with this method. 
It is important to mention that these progenitor detections have not been confirmed yet. 
However, despite the lack of confirmation, SN~2017eaw has nebular spectral analysis and SN~2018aoq has not been analysed via hydrodynamical modelling yet, which make these two SNe relevant in our study for further validation between different methods.
Finally, whenever possible, we compare with the results from late-time spectral modelling found in the literature as well.

This paper can be considered as a companion of a forthcoming paper which analyses a large sample of SNe~II using the same grid of simulations and fitting procedure. In this work we present the stellar evolution calculations we use to obtain the structure of stars at core collapse, the grid of explosion models, the technical part of the fitting procedure and a sanity check.

Our paper is organised as follows. We first present a description of our hydrodynamic code and pre-SN models used in Sect.~\ref{sec:models}. Then, a brief description of the sample of SNe is provided (Sect.~\ref{sec:sample}). In Sect.~\ref{sec:mcmc} we present the fitting procedure we use and its characteristics. In Sect.~\ref{sec:results} we present comparisons between our results with previous studies using different methods. Section~\ref{sec:discussion} provides discussion on LC degeneracies and the limitations of our models, and finally, we summarise our conclusions in Sect.~\ref{sec:conclusions}.

\section{Hydrodynamic simulations} 
\label{sec:models}

\begin{figure}
\centering
	\includegraphics[width=0.48\textwidth]{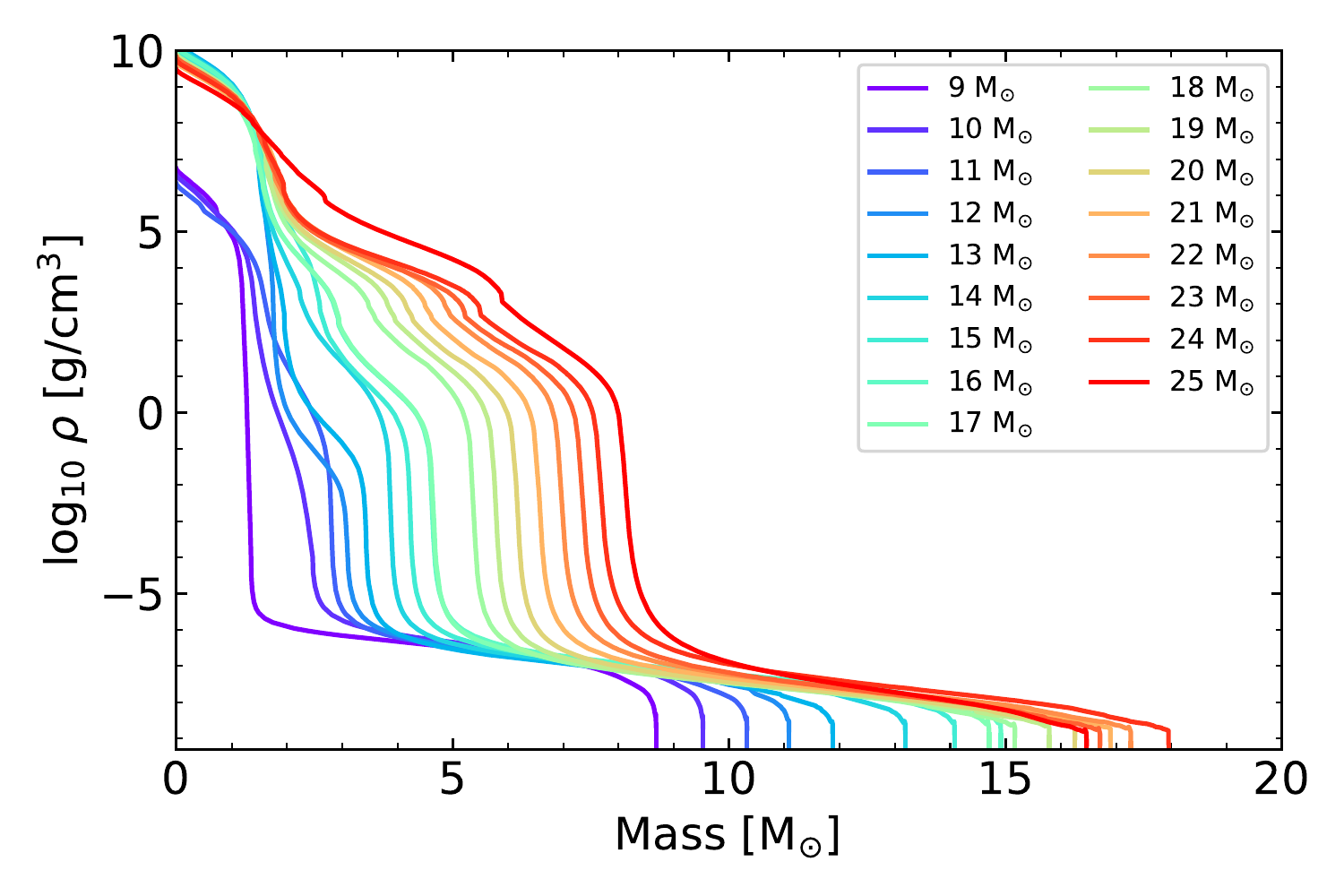}
        \begin{flushleft}
        \caption{Density profiles for the pre-SN models used in this work. The 9, 10, and 11~\ms\ models were calculated up to the end of core carbon burning since the evolution to core collapse for these stars is computationally expensive.}
   		\label{fig:dens_profiles}
		\end{flushleft}
\end{figure}

\begin{figure}
\centering
\includegraphics[width=0.49\textwidth]{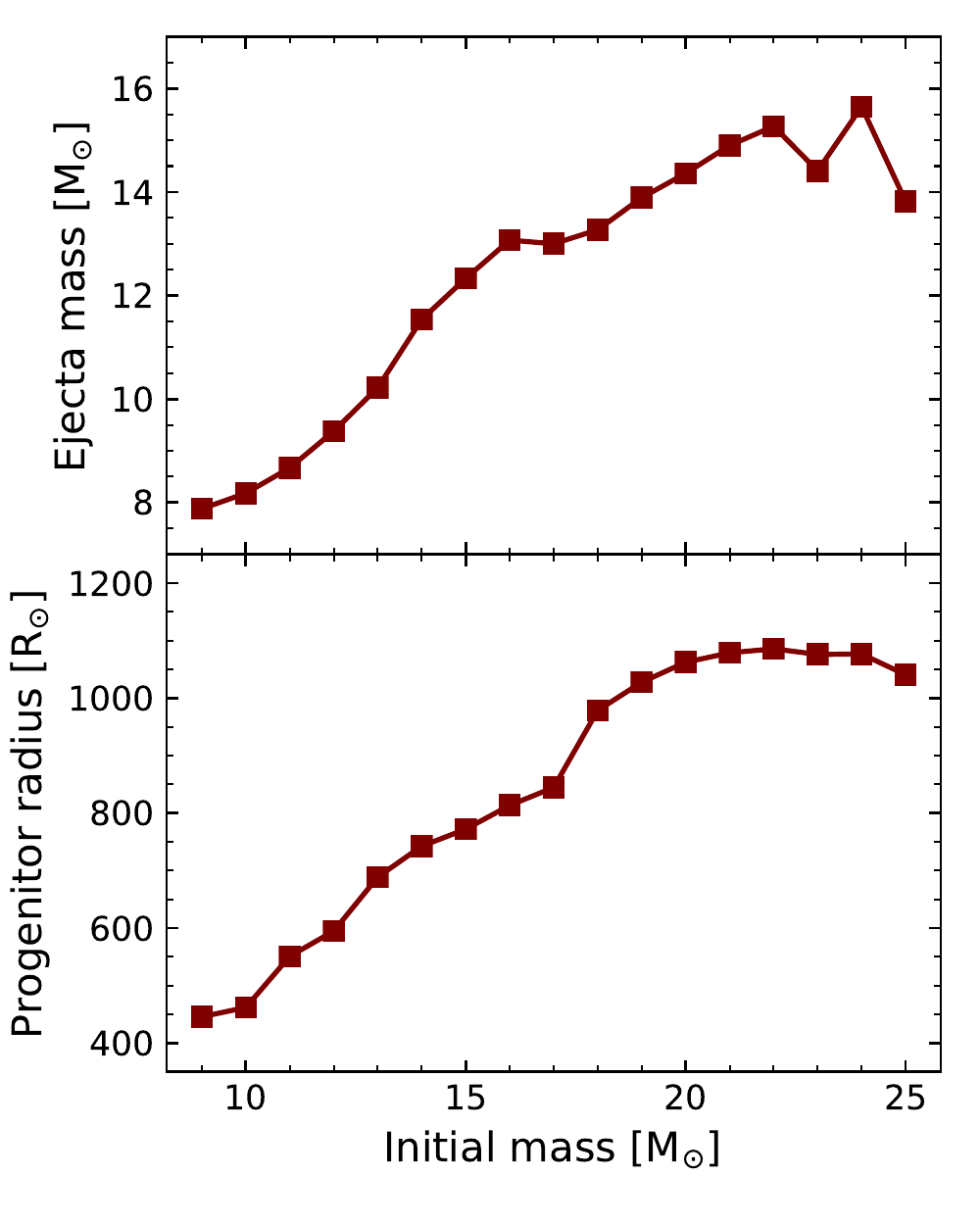}
\caption{Ejecta masses \textit{(top panel)} and progenitor radii \textit{(bottom panel)} for the pre-SN models as a function of the initial mass.}
\label{fig:ejecta}
\end{figure}

\begin{table}
\centering
	\caption{Progenitor properties for the pre-SN models used in this work. $M_{\rm presn}$, $R$, $M_{\rm He}$, and $M_{\rm CO}$ refer to the progenitor final mass and radius, He core size, and CO core size, respectively.}
	\label{table:presn_models}
	\medskip
	\resizebox{0.5\textwidth}{!}{
	\begin{tabular}{c c c c c}
		\hline\hline\noalign{\smallskip}
		\mzams (\ms) & $M_{\rm presn}$ (\ms) & $R$ (\rs) & $M_{\rm He}$ (\ms) & $M_{\rm CO}$ (\ms) \\
		\hline\noalign{\smallskip}
		9  & 8.67 & 445 & 1.33 & 1.19 \\
		10 & 9.53 & 462 & 2.47 & 1.38 \\
		11 & 10.32 & 551 & 2.78 & 1.56 \\
		12 & 11.08 & 594 & 3.05 & 1.75 \\
		13 & 11.87 & 688 & 3.40 & 1.72 \\
		14 & 13.19 & 742 & 3.84 & 2.24 \\
		15 & 14.08 & 772 & 4.18 & 2.51 \\
		16 & 14.92 & 813 & 4.57 & 2.82 \\
		17 & 14.70 & 844 & 4.59 & 2.85 \\
		18 & 15.17 & 978 & 5.30 & 3.43 \\
		19 & 15.79 & 1027 & 5.69 & 3.76 \\
		20 & 16.26 & 1062 & 6.07 & 4.09 \\
		21 & 16.90 & 1078 & 6.46 & 4.44 \\
		22 & 17.27 & 1085 & 6.86 & 4.79 \\
		23 & 16.71 & 1075 & 7.23 & 5.11 \\
		24 & 17.95 & 1076 & 7.57 & 5.42 \\
		25 & 16.47 & 1040 & 8.01 & 5.79 \\
		\hline
	\end{tabular}}
\end{table}

Theoretical LCs are calculated using a 1D Lagrangian hydrodynamical code that simulates the explosion of the SN and produces bolometric LCs and photospheric velocities of SNe \citep{bersten+11}. The explosion is simulated by injecting a certain amount of energy near the centre of the progenitor object, which produces a powerful shock wave that propagates through the star transforming the thermal and kinetic energy of the matter into energy that can be radiated from the stellar surface.

The code assumes that the fluid motion can be described as a 1D, radially symmetric flow and that radiation and matter are strongly coupled, that is to say that local thermodynamical equilibrium (LTE) describes the radiative transfer. The code uses opacity tables calculated assuming LTE and a medium at rest \citep[see][for details]{bersten+11}. 

A pre-supernova model in hydrostatic equilibrium that simulates the conditions of the star before exploding is necessary to initialise the explosion. We use the public stellar evolution code \texttt{MESA\footnote{\url{http://mesa.sourceforge.net/}}} version 10398 \citep{paxton+11,paxton+13,paxton+15,paxton+18,paxton+19} to obtain non-rotating solar-metallicity RSG models.
Each stellar model is evolved from the pre-main sequence until core collapse, which we take as the time when any location inside the stellar model reaches an infall velocity of 1000~km~s$^{-1}$. We use Ledoux criterion for convection and set a mixing-length parameter of $\alpha_{mlt}$~=~2.0, exponential overshooting parameters $f_{\rm ov}$~=~0.004 and $f_{\rm ov,D}$~=~0.001, a semiconvection efficiency $\alpha_{sc}$~=~0.01 according to \citet{farmer+16}, and thermohaline mixing with coefficient $\alpha_{th}$~=~2 \citep{kippenhahn+80}. For every model, we use the “Dutch” wind scheme \citep{dejager+88,vink+01,glebbeek+09} defined in the \texttt{MESA} code with an efficiency $\eta$~=~1. Figure \ref{fig:dens_profiles} shows the density profiles for the 17 pre-SN models used in this work. Some additional properties of the pre-SN models are given in Table~\ref{table:presn_models}. The 9, 10, and 11~\ms\ progenitor models were calculated up to the end of core carbon burning since the evolution to core collapse for these stars is computationally expensive.
The absence of later stages of evolution for these three pre-SN models is noted in Fig.~\ref{fig:dens_profiles} as the inner core density is about three orders of magnitude lower than the rest of the stellar models.
However, only the inner core changes from this part of evolution until core collapse and, additionally, we remove this part for the hydrodynamic calculation as we assume it will collapse and leave a compact remnant. 

The determination of the physical properties of SNe~II is based on describing the bolometric LC and the expansion velocity at the photospheric layers by means of comparing hydrodynamical models with observations.
The morphology of the LC and the evolution of the photospheric velocity are related to physical properties of the progenitor star and the explosion itself, such as the mass ($M_{\rm presn}$) and radius (\emph{R}) prior to explosion, the energy that is transferred to the envelope after core-collapse (denoted as “explosion energy”; \emph{E}), the amount of $^{56}$Ni synthesised in the explosion (\mni) and its degree of mixing into the outer layers of the ejecta (\mix).
Additionally, $M_{\rm presn}$ and \emph{R} depend on the evolution of the star and are directly connected to the progenitor initial mass (\mzams). Figure~\ref{fig:ejecta} shows the ejecta mass (pre-SN mass minus the compact remnant mass) and final radius for every model as a function of \mzams. Ejecta masses cover a range of 7.9$-$15.7~\ms, while progenitor radii are found in the range of 445$-$1085~\rs.

We computed a grid of explosion models in the parameter space covering a \mzams\ range of 9$-$25~\ms\ in intervals of 1~\ms\ (which represent the ranges of ejecta masses and final radius as described above) and explosion energies between 0.1$-$1.5~foe (1~foe~$\equiv$~10$^{51}$~erg) in steps of 0.1~foe with the exception of the largest masses and lowest energies due to numerical difficulties. 
For the 20~\ms\ and 21~\ms\ models, the lowest explosion energies are 0.2~foe and 0.3~foe, respectively. Models of 22~\ms\ and 23~\ms\ were calculated for explosion energies higher than 0.4~foe, and for the 24~\ms\ and 25~\ms\ models, only explosion models with energies higher than 0.5~foe are available.
We also consider \mni\ in the range of 0.01$-$0.08~\ms\ in intervals of 0.01~\ms, together with nickel masses of 0.0001 and 0.005~\ms\ to be consistent with the lowest SN~II estimated \Ni\ masses in the literature \citep{muller+17,anderson19}. To account for the effect of the spatial distribution of \Ni\ within the ejecta we consider three degrees of \mix\ for each model: out to the 20\%, 50\% and 80\% of the pre-SN structure in mass coordinate.  
 
Recent studies have shown that the interaction of the ejecta with a circumstellar material (CSM) shell surrounding the progenitor star can affect the early LC of SNe \citep{gonzalez+15,moriya+17,forster+18,morozova+18}. In the current work, we do not attempt to characterise the CSM. In this way, we do not include any CSM surrounding the star in our explosion models and focus on deriving intrinsic properties of the progenitors. Therefore, we restrict the analysis of the observed LC to times later than 30~days after explosion (see Sect.~\ref{sec:results}).

\section{Data sample} 
\label{sec:sample}

\begin{figure}
\centering
	\includegraphics[width=0.5\textwidth]{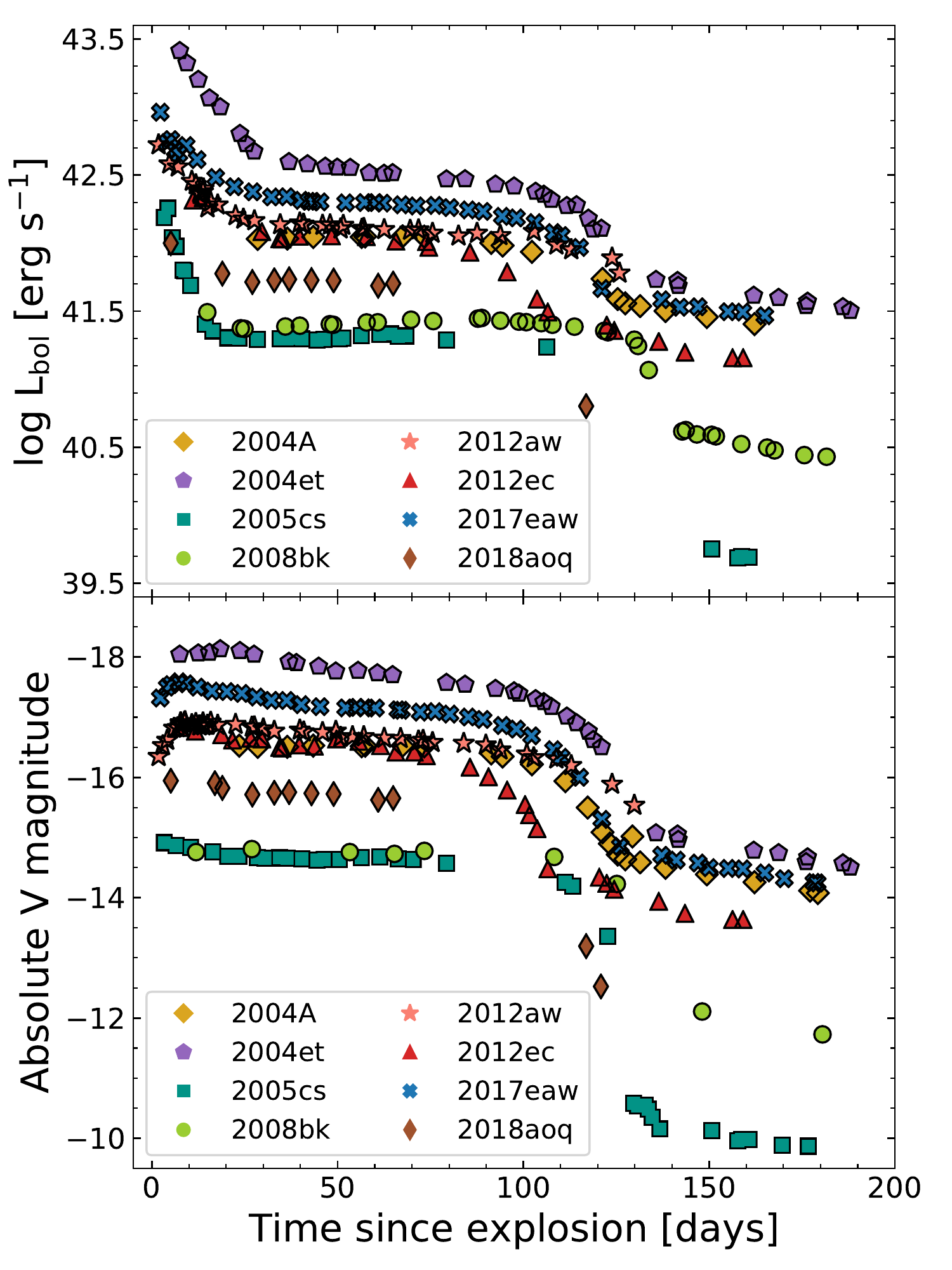}
        \begin{flushleft}
        \caption{Bolometric LCs (top panel) and absolute $V$-band LCs (bottom panel) of SN~2017eaw and SN~2018aoq in comparison with the well-studied SN~2004A, SN~2004et, SN~2005cs, SN 2008bk, SN~2012ec, and SN 2012aw. Absolute curves were computed using the distances and reddenings in Table~1 of MB19, except SN~2004et for which we use the new estimates of the distance (see text).}
   		\label{fig:comparacion}
		\end{flushleft}
\end{figure}

In this work, we aim to study if our method to infer physical properties of SNe~II is consistent with other methods in the literature. For this purpose we use some of the most well-studied SNe~II. In particular, we use the same sample analysed in MB19 defined as SNe~II that: a) have well-constrained pre-explosion progenitor detections, b) have post-explosion images confirming the disappearance of the progenitor, and c) have sufficiently well sampled photospheric-phase observations to enable accurate modelling fits.
In addition, we have included two objects to the sample, SN~2017eaw and SN~2018aoq, as these are the latest published SNe~II with progenitor identification in pre-explosion images. However, these identifications have not been confirmed yet by post-explosion images. Information for these two SNe are presented in Sects~\ref{sec:17eaw} and \ref{sec:18aoq}. Details for the other objects can be found in MB19 and references therein.

Following our sample definition, we calculate their bolometric LCs using the correlation between bolometric correction and colours inferred in \citet{bersten+09}, which allows to calculate bolometric luminosities using only two optical filters. We use the dispersion values listed in Table~1 of \citet{bersten+09} and the uncertainties in colours to estimate the uncertainty in the bolometric luminosities via error propagation. Neither the uncertainties in the distance nor extinction were propagated to the bolometric LCs as we include an additional parameter in the MCMC procedure that models it (Sect.~\ref{sec:mcmc}). The values of host extinctions, distances and explosion times are the same that those presented in Table~1 of MB19 with the exception of the distance assumed for SN~2004et. Here we recalculate the bolometric LC of SN~2004et using the most recent estimation for the distance to its host galaxy NGC~6946 (see Sect.~\ref{sec:17eaw}).

In addition to the bolometric luminosity, an estimation of the photospheric velocity is also needed to compare with the models. This velocity can be estimated through the measurement of certain spectroscopic lines. We use the \ion{Fe}{ii} $\lambda$5169~\AA\ line since this line is formed in internal regions of SNe and it has been proposed as a good estimator of the photospheric velocity \citep[see][and Sect.~\ref{sec:limitations} for discussion]{dessart+05}.

With the aim of assessing the nature of progenitors and comparing our constraints with results obtained from pre-explosion information and late-time spectral modelling, we set out to perform a detailed modelling of the available observations.
Figure~\ref{fig:comparacion} shows the bolometric and absolute $V$-band LCs of SN~2017eaw and SN~2018aoq in comparison to the sample from MB19. It seems like SN~2017eaw is intermediate in luminosity between, for example, SN~2012aw and the more luminous SN~2004et, while SN~2018aoq appears to be an intermediate case between the normal and the low-luminosity SNe~II \citep{oneill+19}. From this comparison we also note that the last two $V$-band observations of SN~2018aoq constrain the end of the plateau phase.

\subsection{SN~2017eaw}
\label{sec:17eaw}

SN~2017eaw was discovered on 2017 May 14.238 UT in NGC~6946 at an unfiltered magnitude of 12.8~mag \citep{wiggins17}. This object was classified as a young SN~II by \citet{cheng+17}, \citet{xiang+17}, and \citet{tomasella+17}. \citet{wiggins17} also observed the site of explosion on 2017 May 12.20 UT but nothing was visible. This limitation in the detection restrict the uncertainty in the explosion epoch (\te) to only one day. We adopt \te\ as JD~2457886.72~$\pm$~1.01 \citep{rui+19}.

Pre-explosion images of the SN location were obtained with the Hubble Space Telescope (HST) and the Spitzer Space Telescope covering the last $\sim$13~yr before explosion. The progenitor can be detected in eight optical and infrared bands making it one of the most-characterised progenitors to date. Several works studying the progenitor properties have been published. \citet{kilpatrick+18} found a RSG progenitor compatible with an initial mass of $\sim$13~\ms. They detected an increase in its 4.5~$\mu m$ luminosity over the final 3~yr before the explosion and argue that it is a signature of circumstellar dust near the progenitor star. Moreover, \citet{rui+19} found a narrow and blueshifted H$\alpha$ emission component in a spectrum taken a few hours after discovery that disappeared in less than two days, suggesting the presence of a CSM shell. The authors propose that the progenitor could have experienced a dramatically enhanced mass-loss during the last 1--2~yr before explosion. They also found a RSG progenitor with an initial mass of 12~$\pm$~2~\ms.
On the other hand, \citet{vandyk+19} established that the progenitor was a dusty, luminous RSG consistent with an initial mass of $\sim$15~\ms.
Unfortunately, these three studies had assumed different values for the distance to the object.
Recently, \citet{murphy+18}, \citet{anand+18}, and \citet{vandyk+19} used archival HST data taken in the outer regions of NGC~6946 to measure the tip of the red giant branch and infer the distance to the galaxy. All these studies arrive at the same value for the distance, to within the uncertainties.
In addition, \citet{eldridge+19a} used these new measurements to re-evaluate the final luminosity for some SN progenitors in NGC~6946 (SN~2004et, among others). With the new distance they estimate that the initial mass of the progenitor of SN~2017eaw is 14$^{+3.0}_{-3.5}$~\ms, consistent with the results from \citet{vandyk+19} for the same distance. In this paper we set the distance in 7.73~$\pm$~0.78~Mpc and the total extinction as 0.941~mag \citep[see discussion in][]{vandyk+19}.

Additionally, using nebular-phase spectral modelling, \citet{vandyk+19} and \citet{szalai+19} analysed late-time spectra of SN~2017eaw and found that a progenitor with initial mass near 15~\ms\ is most consistent with observations.

In this study, we take photometric data and \ion{Fe}{ii} $\lambda$5169~\AA\ line velocities from \citet{szalai+19}. Additional optical photometry can also be found in \citet{tsvetkov+18}, \citet{vandyk+19}, \citet{rui+19}, and \citet{buta+19}.

\subsection{SN~2018aoq}
\label{sec:18aoq}

SN~2018aoq was discovered on 2018 April 01.43 in the galaxy NGC 4151 by the Lick Observatory Supernova Search (LOSS) at the unfiltered magnitude of 15.3 mag \citep{nazarov+18}. \citet{yamanaka18} carried out spectroscopic observations on 2018 April 02 using the Hiroshima One-shot Wide-field Polarimetry (HOWPol) installed to the 1.5-m Kanata telescope and found a spectrum dominated by a blue continuum and the H$\alpha$ line with P-Cygni profile consistent with a young SN~II. \citet{oneill+19} presented optical imaging data and spectra using a combination of Asteroid Terrestrial-impact Last Alert System \citep[ATLAS,][]{tonry+18,smith+20}, the 2.0m Liverpool Telescope (LT) and the 2.5m Nordic Optical Telescope (NOT) as part of the NOT Unbiased Transient Survey (NUTS). They also presented an estimation of the total reddening and a new estimation of the distance that is in good agreement with the value based on geometric methods. In addition, from the ATLAS non-detection on 2018 March 28, the explosion epoch is well constrained to within four days. We adopt the same values of distance ($d$~=~18.2~$\pm$~1.2~Mpc), reddening ($E(B-V)_{\rm tot}$ = 0.04~mag), and explosion epoch (JD~2458208.5) as \citet{oneill+19} in what follows.
 
Archival pre-explosion images of the SN site are available. These images were taken with the Wide Field Camera 3 (WFC3) on board the HST approximately 2~yr before explosion. From these observations, \citet{oneill+19} detected a source at the SN location in four bands: F350LP, F555W, F814W and F160W. From fits to the spectral energy distribution of the progenitor candidate, they found a luminosity range of log($L/L_{\odot}$)~=~4.56$-$4.83 and an effective temperature of $T$~=~3500~$\pm$~150~K, implying an M-type red supergiant progenitor. Using single and binary star models, they conclude that the explosion of a star with a ZAMS mass of 10~$\pm$~2~\ms\ is the most favoured scenario. 
\section{Fitting procedure}
\label{sec:mcmc}

\begin{figure}
\centering
\includegraphics[width=0.5\textwidth]{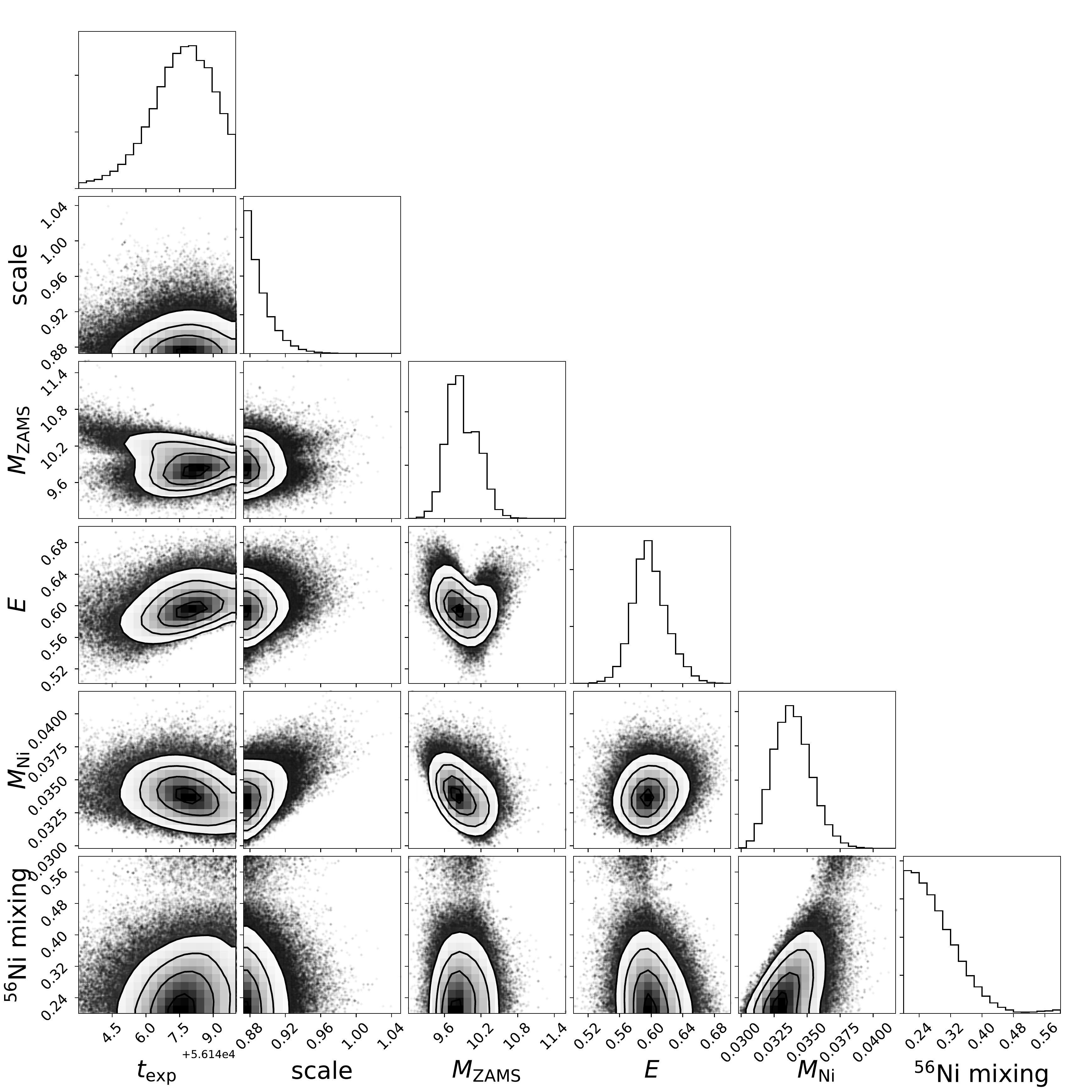}
\caption{Corner plot of the joint posterior probability distribution of the parameters for SN~2012ec.}
\label{fig:corner_12ec}
\end{figure}

In MB19, the authors derive physical parameters for the sample presented in Sect.~\ref{sec:sample}, with the exception of SN~2017eaw and SN~2018aoq, based on visual comparisons between observations and hydrodynamic simulations. In this work we aim to develop and test a robust fitting procedure that automatically obtains optimal solutions for the fitted parameters in a statistically sound manner, following \citet{forster+18}. Thus, we computed a large grid of models of bolometric LCs and photospheric velocity evolution. The range of physical parameters considered is described in Sect.~\ref{sec:models}. However, these models may not be enough when trying to fit SN observations using statistical inference techniques. We need to be able to interpolate between models with different physical parameters. 
First we define the set of parameters for which we want to compute an interpolated model, that is, \mzams, \emph{E}, \mni, and \mix. Then, in our grid of simulations we find the models with the closest values in all the physical parameters. The interpolated LC and velocity evolution are calculated using Eqs.~2 and 3 of \citet{forster+18}. This is a robust and quick method which attempts to provide a scale--free interpolation, relevant when combining variables with different physical dimensions. It also allows for irregular grids of models in the space of parameters to be used.
We show examples of interpolated models in Fig.~\ref{fig:interpolation}.

Having a powerful interpolation method we can attempt to infer the physical parameters using Bayesian statistics, i.e. computing the posterior probability of the model parameters given the observations and assuming prior distributions.
To do this we use a MCMC sampler which uses an affine invariant approach \citep{goodman+10}. This method estimates the properties of a distribution by examining random samples from the distribution which are generated using parallel Markov chains. The characteristic of the Markov chains is that each random sample is used to generate the next random sample. While each new sample depends on the one before it, new samples do not depend on any samples before the previous one \citep{ravenzwaaij+18}. This method is implemented via \texttt{emcee} in python \citep{emcee}.

We run the MCMC sampler using flat distributions as priors for the following parameters: \te, \mzams, \emph{E}, \mni, and \mix, allowing the sampler run within the observational uncertainty of the \te. In addition, we define a variable named \emph{scale} for which we use a Gaussian prior.
The \emph{scale} parameter multiplies the bolometric luminosity by a constant factor to allow for errors in the bolometric LC due to the uncertainties in the distance and extinction. For this parameter we use a Gaussian prior centred at 1.0 and with a standard deviation equal to the uncertainties in the distance estimation. It is worth emphasising that the uncertainties in the distance generally dominate over those of extinction. In this way, we assume that the distance errors include the extinction errors. Additionally, we constrained the \emph{scale} parameter to $\pm$1$\sigma$.
The reason for this choice is as follows. Progenitor initial mass is not an observational parameter. For example, in pre-SN imaging, the detection of the progenitor star gives the observed luminosity of the object close to the explosion epoch, which is then converted to an initial mass using stellar evolution calculations. Thus, the initial mass depends on the luminosity of the progenitor, and this depends on the distance to the object, host galaxy extinction, among others.
Constraining the \emph{scale} parameter to $\pm$1$\sigma$ we make sure that the physical properties we derive are consistent with the distance and extinction estimates.
In this context we can compare our mass estimation with that from pre-SN imaging (or any other method) as we assume the same range of distances and total extinction.

We use 400 parallel samplers (or walkers) and 10000 steps per sampler, with a burn-in period of 1000 steps. These numbers were set via trial and error through checking randomness and stationarity of the chains. The walkers are randomly initialised covering the entire parameter space. 

An example corner plot with the posterior probability distributions can be visualised in Fig.~\ref{fig:corner_12ec}. It is seen that the marginal distributions for the \te\, \emph{scale}, and \mix\ are limited.
The posterior distribution marginalised over the \mix\ parameter is restricted to values above 0.2 due to limitations in our model. Additionally, the \te\ and the $scale$ are constrained to the uncertainties in the explosion epoch and distance, respectively. If we relax the priors for these two parameters we obtain a similar marginal distribution of the physical parameters (see Fig.~\ref{fig:corner_12ec_texp_scale}). 
The corner plots for the entire sample are shown in Appendix~\ref{ap:corner}, and the models drawn from the posterior distribution for the SN sample are shown in Figs.~\ref{fig:models1} and \ref{fig:models2}.
Examples of autocorrelation plots and trace plots are presented in Figs.~\ref{fig:autocorrelation} and \ref{fig:traceplot}, respectively.

\section{Results}
\label{sec:results}

\begin{figure*}
\centering
\includegraphics[width=0.89\textwidth]{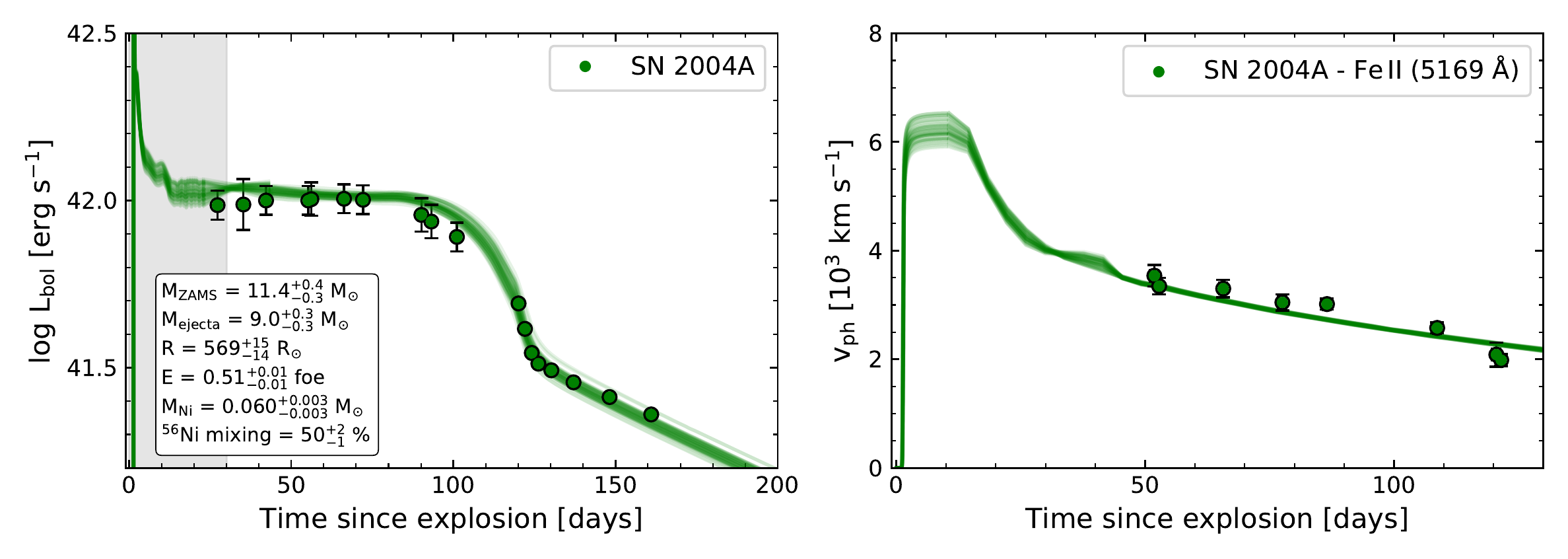}
\includegraphics[width=0.89\textwidth]{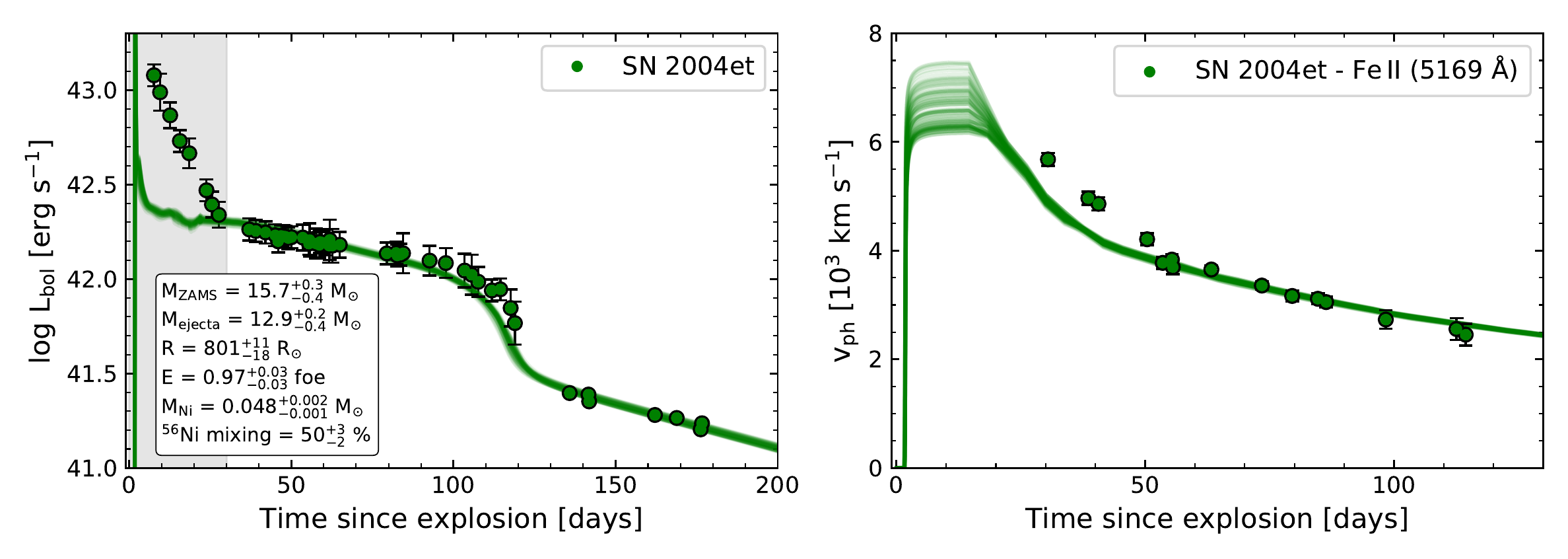}
\includegraphics[width=0.89\textwidth]{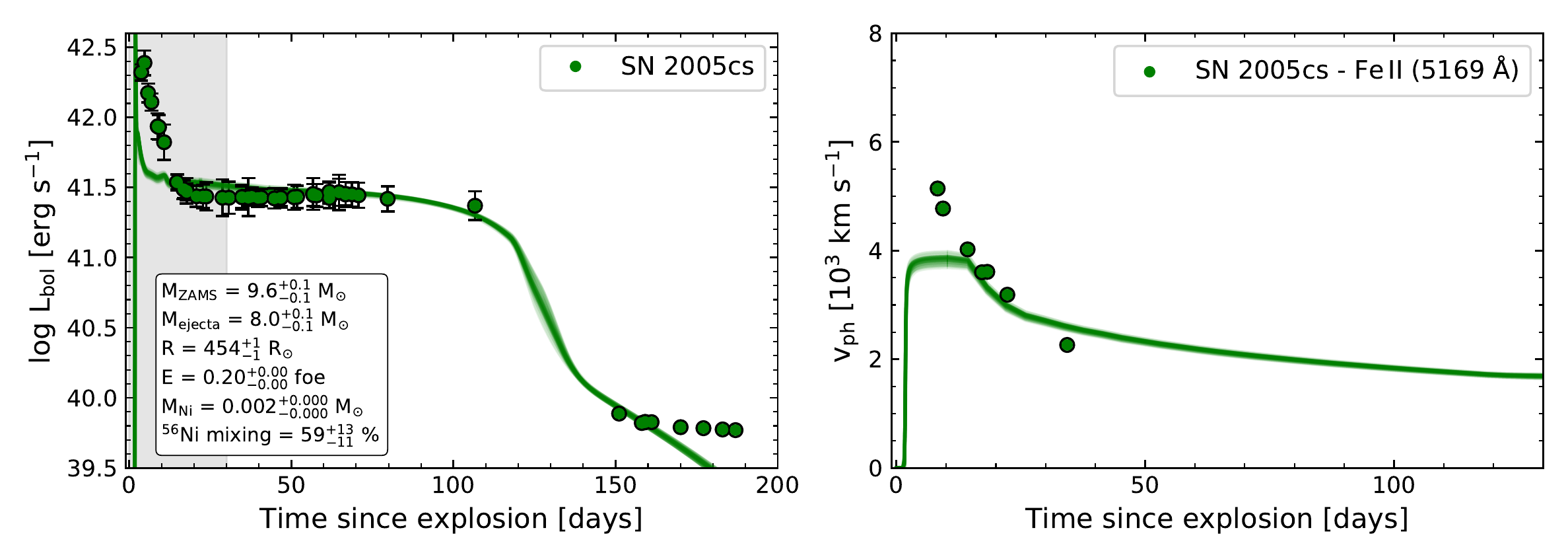}
\includegraphics[width=0.89\textwidth]{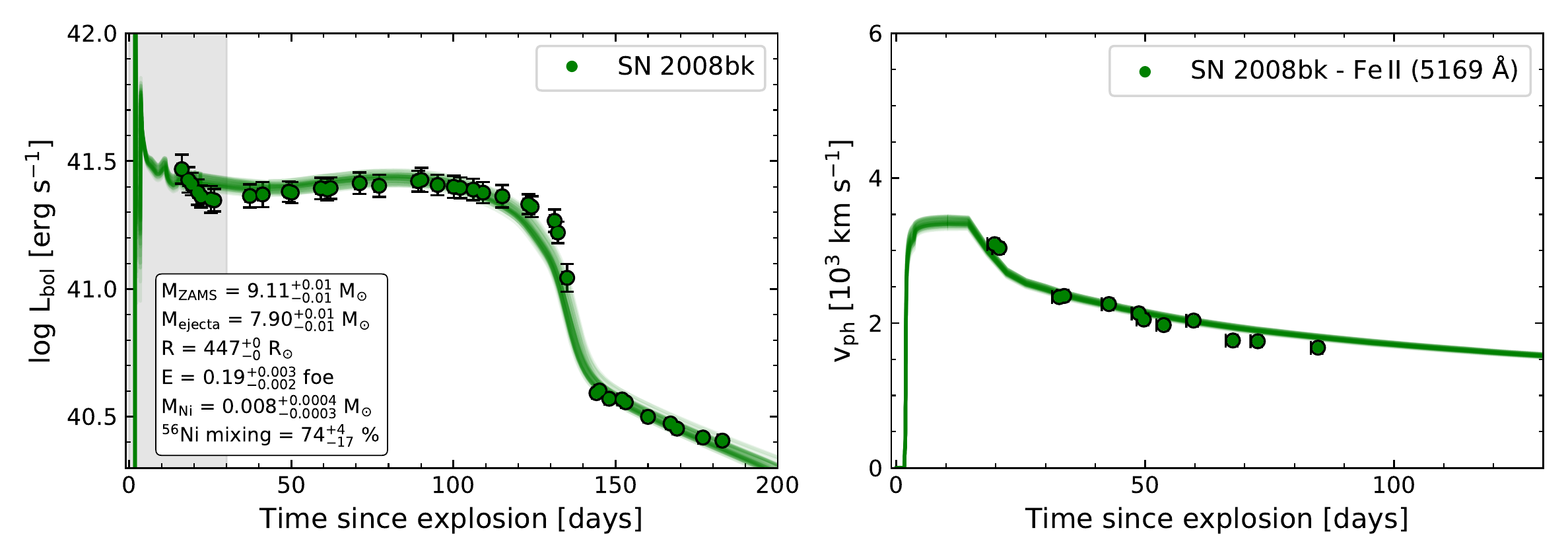}
\caption{Comparison between models (solid lines) and observations (filled dots) for our SN sample. We show 50 models randomly chosen from the posterior probability distribution. \textit{Left:} bolometric LC. \textit{Right:} evolution of the photospheric velocity. \textit{From top to bottom:} SN~2004A, SN~2004et, SN~2005cs, and SN~2008bk. The grey shaded region shows the early data we removed from the fitting. For SN~2004et we show the results using $d$~=~5.9~$\pm$~0.4~Mpc to calculate the bolometric LC.}
\label{fig:models1}
\end{figure*}

\begin{figure*}
\centering
\includegraphics[width=0.89\textwidth]{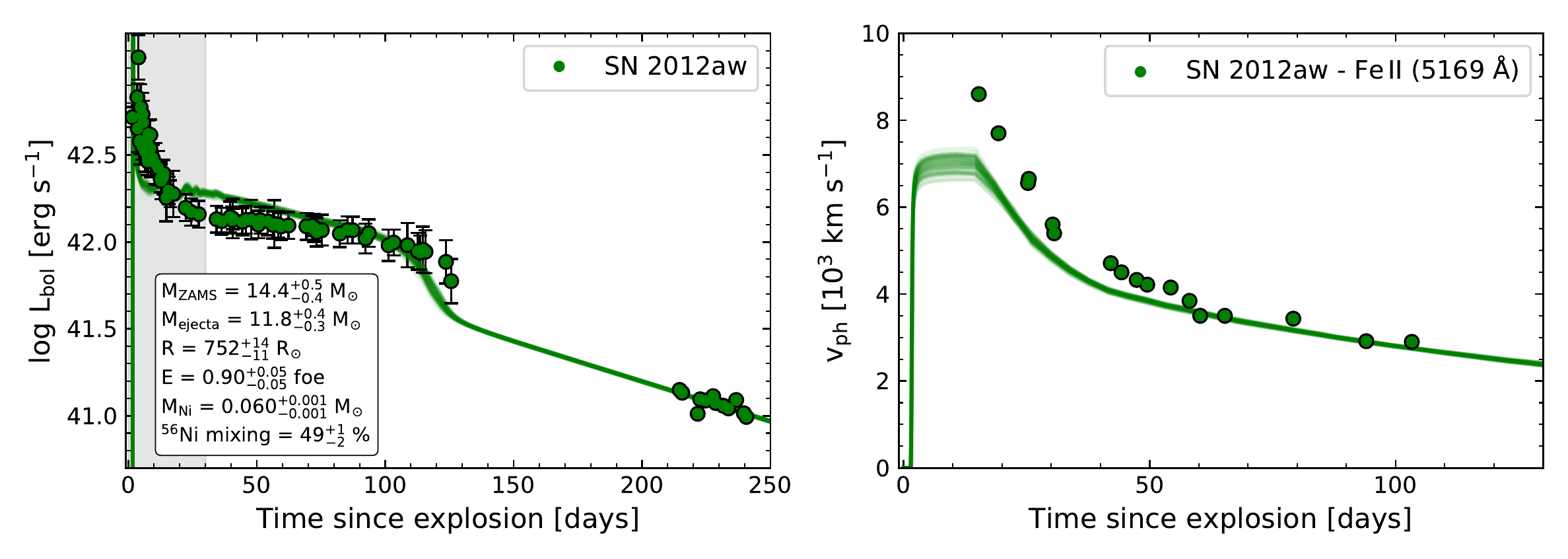}
\includegraphics[width=0.89\textwidth]{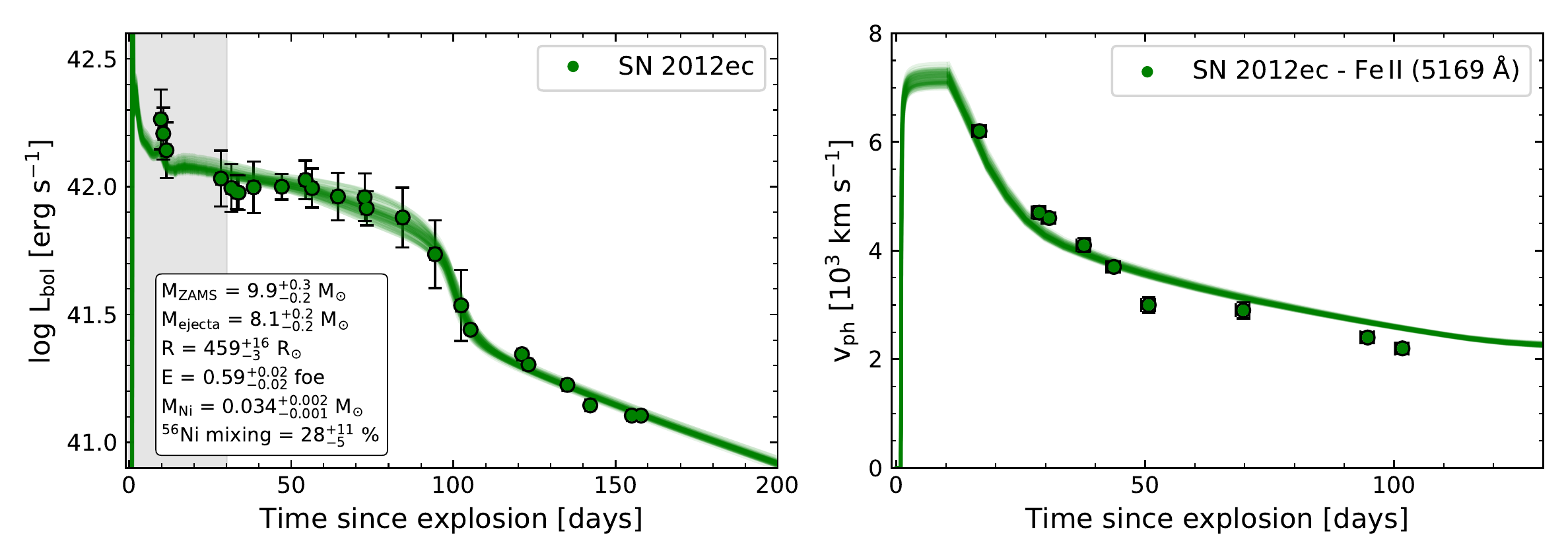}
\includegraphics[width=0.89\textwidth]{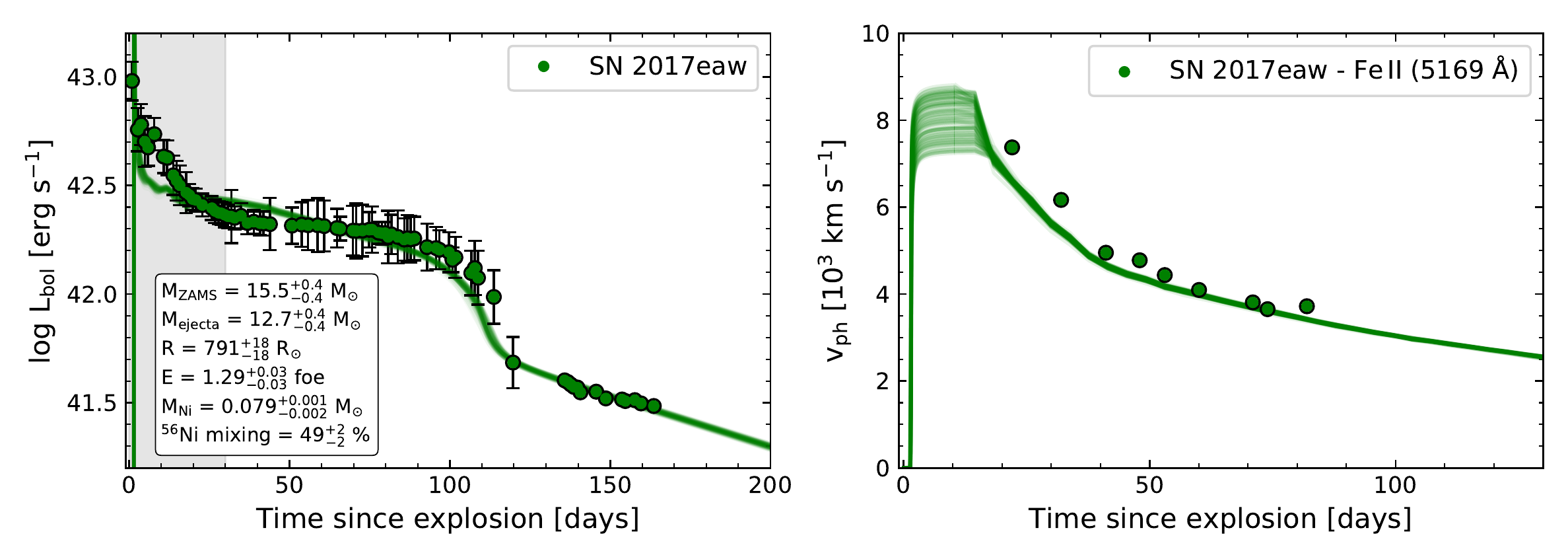}
\includegraphics[width=0.89\textwidth]{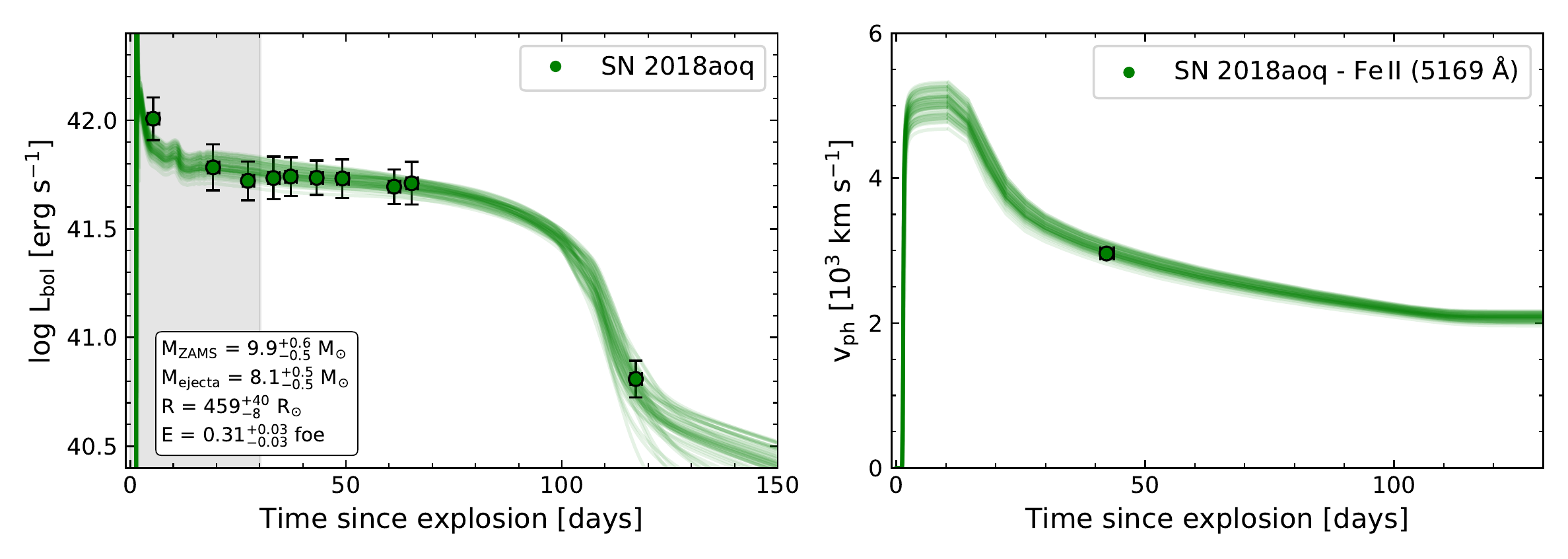}
\caption{Comparison between models (solid lines) and observations (filled dots) for our SN sample. We show 50 models randomly chosen from the posterior probability distribution. \textit{Left:} bolometric LC. \textit{Right:} evolution of the photospheric velocity. \textit{From top to bottom:} SN~2012aw, SN~2012ec, SN~2017eaw, and SN~2018aoq. The grey shaded region shows the early data we removed from the fitting.}
\label{fig:models2}
\end{figure*}

\begin{table*}
\caption{Physical parameters derived from the hydrodynamic modelling using MCMC methods. We characterise the results by the median of the posterior distribution and the 16th and 84th percentiles as our lower and upper uncertainties.}
\label{table:results}
\centering                
\begin{tabular}{c c c c c c c}      
\hline\hline\noalign{\smallskip}    
SN & \te~[MJD] & scale & \mzams~[\ms] & $E$~[foe] & \mni~[\ms] & \Ni\ mixing~[\%] \\
\hline\noalign{\smallskip}       
2004A & 53009.2$^{+0.8}_{-0.9}$ & 0.90 $\pm$ 0.04 & 11.43$^{+0.35}_{-0.34}$ & 0.51 $\pm$ 0.01 & 0.060$^{+0.004}_{-0.003}$ & 51$^{+3}_{-2}$ \\
\noalign{\smallskip}
2004et & 53270.0$^{+0.1}_{-0.2}$ & 0.88$^{+0.03}_{-0.02}$ & 15.72$^{+0.30}_{-0.44}$ & 0.97 $\pm$ 0.03 & 0.048$^{+0.002}_{-0.001}$ & 50$^{+3}_{-2}$ \\
\noalign{\smallskip}
2005cs & 53548.6$^{+0.6}_{-0.7}$ & 1.36 $\pm$ 0.01 & 9.55 $\pm$ 0.09 & 0.200$^{+0.002}_{-0.001}$ & 0.002 $\pm$ 0.0002 & 60$^{+13}_{-11}$ \\
\noalign{\smallskip}
2008bk & 54539.7$^{+0.9}_{-1.5}$ & 0.95$^{+0.03}_{-0.01}$ & 9.11$^{+0.01}_{-0.01}$ & 0.190 $\pm$ 0.003 & 0.008 $\pm$ 0.0004 & 74$^{+4}_{-17}$ \\
\noalign{\smallskip}
2012aw & 56002.8$^{+0.1}_{-0.2}$ & 0.99 $\pm$ 0.01 & 14.35$^{+0.50}_{-0.37}$ & 0.90 $\pm$ 0.05 & 0.060 $\pm$ 0.001 & 49$^{+2}_{-3}$ \\
\noalign{\smallskip}
2012ec & 56147.8$^{+1.3}_{-1.5}$ & 0.89$^{+0.02}_{-0.01}$ & 9.87$^{+0.29}_{-0.21}$ & 0.59 $\pm$ 0.02 & 0.034$^{+0.002}_{-0.001}$ & 28$^{+12}_{-6}$ \\
\noalign{\smallskip}
2017eaw & 57887.1$^{+0.1}_{-0.2}$ & 1.04 $\pm$ 0.02 & 15.47$^{+0.45}_{-0.43}$ & 1.29 $\pm$ 0.03 & 0.079$^{+0.001}_{-0.002}$ & 50 $\pm$ 2 \\
\noalign{\smallskip}
2018aoq & 58207.8$^{+1.5}_{-1.2}$ & 1.02$^{+0.08}_{-0.09}$ & 9.87$^{+0.57}_{-0.55}$ & 0.31 $\pm$ 0.03 & --- & --- \\
\noalign{\smallskip}
\hline
\end{tabular}
\begin{flushleft}
\tablefoot{Results for SN~2004et correspond to $d$ = 5.9 $\pm$ 0.4~Mpc}
\end{flushleft}
\end{table*}

We have developed a fitting procedure to derive physical parameters of SNe~II from the hydrodynamic modelling of LCs and photospheric velocities using the models described in Sect.~\ref{sec:models} and the method from Sect.~\ref{sec:mcmc}.

It is known that the presence of a dense CSM affect the early evolution of SNe~II with almost no effect at later epochs \citep[t $\gtrsim$ 30 days after explosion, see Figs.~3 and 4 of][]{morozova+18} where the evolution is dominated by the hydrogen recombination and radioactive decay. This is correct if the ejecta interacts with a low-mass CSM (Englert in prep.). 
The effect of the ejecta-CSM interaction can dominate the behaviour of the early LC suggesting that the general characteristics of the CSM have a significant role in this early phase, and not the progenitor properties as mentioned by \citet{utrobin+08}.
As we have noticed in Sect.~\ref{sec:models}, we did not include CSM in our set of progenitor models as we are interested in deriving global properties of the SN progenitor instead of analysing the CSM properties. Therefore, we do not consider the first 30~days of evolution of the observed LC in our fitting procedure. 
As a consequence, it is expected to have some differences between the models and the observations during the cooling phase, which is strongly affected by CSM interaction (where it exists).
Despite this, we decided not to remove the early data from the velocity evolution. If we do not take into account these observations, the discrepancy between the fitted models and the observed velocity evolution at early times could be large.
This could result in incompatibilities as the interaction of the ejecta with a CSM produces a decrease of the photosperic velocities at early times \citep[see, e.g. Fig.~13 of][]{rodriguez+19}. Then, our velocity models should be of the same order of magnitude or with higher velocities than early observations.

Figures \ref{fig:models1} and \ref{fig:models2} show models drawn from the posterior distribution for each SN in our sample, together with median and 1$\sigma$ confidence range for every parameter. The ejecta mass and the progenitor radius were interpolated linearly to the \mzams\ we derived from the fitting. We also report the results in Table~\ref{table:results}. It is important to note that the published errors on our estimated progenitor and explosion parameters are statistical in nature. The size of the errors indicate that out fitting technique is robust. However, these errors do not take into account systematics such as the uncertainties in stellar evolution modelling. There are a number of additional parameters which one could change in MESA that would give different pre-SN configurations (for the same \mzams). A full exploration of these effects is beyond the scope of this work and will be the focus of future efforts. Here we have used `standard' values for various stellar evolution parameters, which may not cover the full parameter space. As a consequence, the errors on the physical parameters are likely to be underestimated.

We can see good quality fits for the whole sample. However, we note some issues.
First, it is important to note that SN~2018aoq presents only a few observations for the LC which includes the photospheric phase and the transition to the radioactive tail phase (see Fig.~\ref{fig:comparacion}), and only one measurement of the \ion{Fe}{ii} 5169~\AA\ velocity. Since there are no observations during the radioactive tail phase, the amount of \Ni\ synthesised cannot be derived \citep{hamuy03}.

We were not able to obtain results for SN~2004et. LCs drawn from the posterior distribution do not represent the observed LC. While the evolution of the photospheric velocity and the radioactive tail phase are well reproduced, the maximum a posteriori (MAP) model is fainter by an average of 0.27~dex during the photospheric phase. 
With the most recent estimation for the distance, the bolometric LC becomes remarkably luminous. Additionally, this SN presents an extensive plateau phase ($\sim$125 days). The combination of both properties makes this SN an outlier in our sample and, unfortunately, we could not find any model in our grid that matches these properties (see Fig.~\ref{fig:sn2004et_badfit}). 
As we mentioned in Sect.~\ref{sec:mcmc}, we constrain the $scale$ parameter to the uncertainties in the distance estimates. Since with this restriction we could not find any set of parameters that fit the observations of this object, we test if solutions were possible in which the \emph{scale} parameter is unconstrained. In this context we find a solution, but with a \emph{scale} factor that represents a shorter distance to the SN which is $\sim$0.4 times the current distance estimate, or a combination of shorter distance and lower extinction. In both cases this distance is outside the range of allowed distances by the latest estimates.
These results may indicate that the shorter distance is in fact correct or that our modelling cannot reproduce the observations because our range of initial models and explosion energies does not include any model compatible with this object.
Recently, three studies arrive at similar distance estimates for its host galaxy \citep{murphy+18,anand+18,vandyk+19}. Hence, there may be an indication of the necessity to change some parameters in the evolutionary calculations to obtain a different progenitor structure that reproduces the properties of this SN. 
Therefore, we discard this solution for SN~2004et as it is not compatible with the new estimates of the distance.
However, as we want to compare the results of this work using hydrodynamic simulations and MCMC fitting  with those using other methods to look for compatibility, we modelled SN~2004et again but now assuming a shorter distance of 5.6~$\pm$~0.4~Mpc \citep{smartt+09} as this is the previous estimate of the distance. The best fitting models for SN 2004et with this distance are shown in the second panel of Fig.~\ref{fig:models1}.

Additionally, we find a poor agreement in the radioactive tail phase for SN~2005cs. It seems that the decline rate of our model does not follow the observations. 
\citet{pastorello+09} already pointed out that the decline rates of SN~2005cs in different bands are significantly smaller than the decline rate expected from the $^{56}$Co decay and describe this phenomenon as a residual contribution from radiation energy, as first suggested by \citet{utrobin07}. According to \citet{utrobin07}, by the end of the optically thick phase the total radiation energy is not exhausted completely. A radiation flow generated in the warmer inner layers propagates throughout the optically thin layers and results in an additional source of energy. This transitional phase \citep[labelled as \emph{plateau tail phase} by][]{utrobin07} was also observed in SN~1999em and SN~1999eu. 

Finally, we also find discrepancies in the velocities of SN~2005cs and SN~2012aw at early times. The largest difference found is about 1000~km~s$^{-1}$.

\subsection{Comparison with the analysis of pre-explosion imaging}
\label{sec:preexp}

\begin{figure}
\centering
\includegraphics[width=0.5\textwidth]{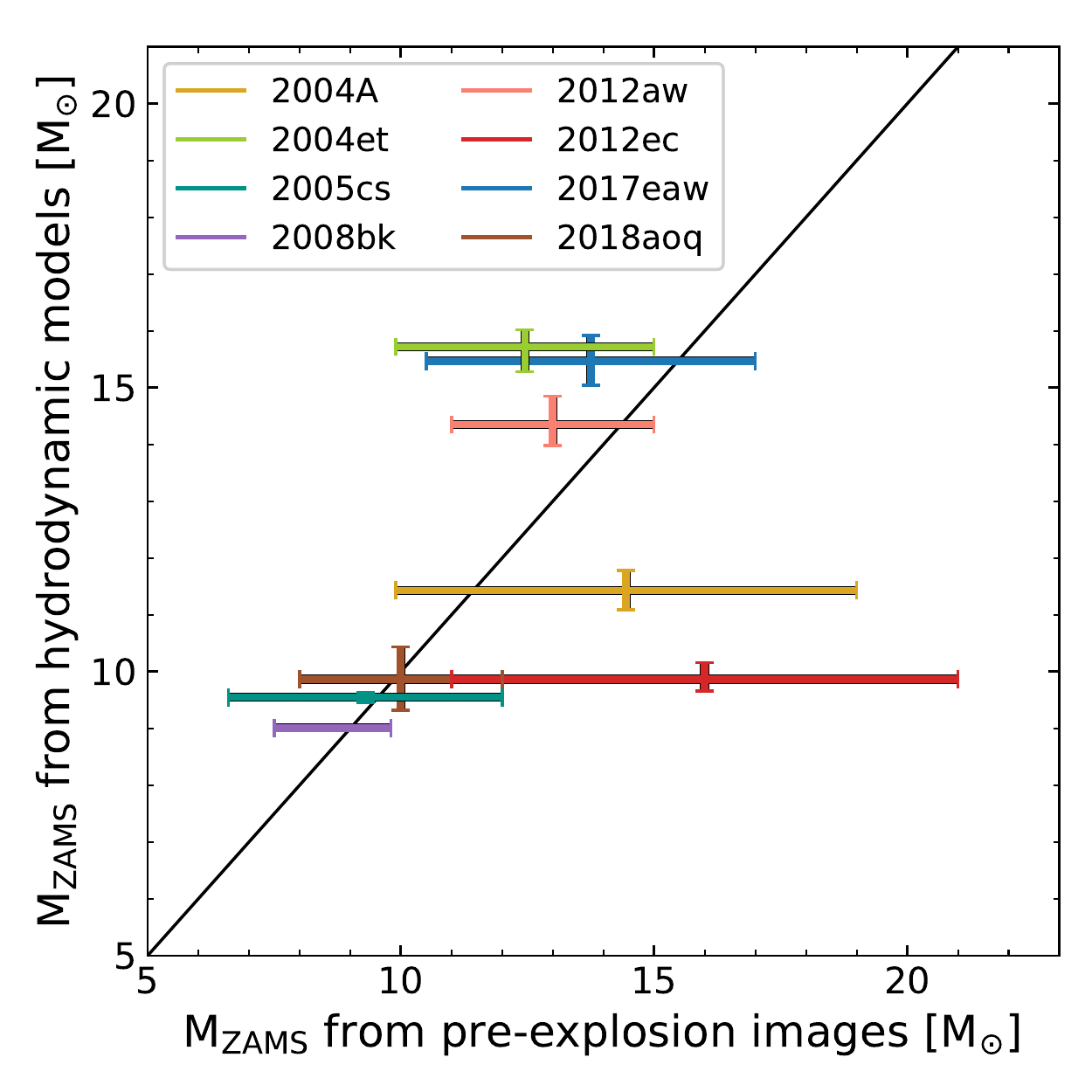}
\caption{Comparison between \mzams\ obtained in this work using hydrodynamic simulations and MCMC methods and those based on the analysis of progenitors in pre-explosion images. For cases where multiple values of \mzams\ from pre-SN imaging exist, we take the complete range of values predicted in the literature instead of using some specific value and its uncertainties. In these cases, the central point of the initial mass is the midpoint of the range.}
\label{fig:hydro_preexp}
\end{figure}

Here we compare our estimation of the progenitor initial mass to those derived from the direct analysis of the progenitor star in pre-explosion images.
We use the results from \citet{smartt15} and \citet{davies+18} for SN~2004A, SN~2004et, SN~2005cs, and SN~2012ec. It is important to remark that these two papers use the old estimate for the distance to SN~2004et so we can compare with our results.
For SN~2008bk we use a combination of the results from \citet{mattila+08}, \citet{smartt+09}, \citet{vandyk+12a}, and \citet{davies+18}. We discard the results from \citet{maund+14b} as their mass determination is subject to a foreground extinction almost ten times higher than the later estimation made by \citet{maund17}.
The solutions of \citet{kochanek+12}, \citet{smartt15}, and \citet{davies+18} were used for SN~2012aw \citep[see discussion in][]{smartt15}.
For SN~2017eaw we make use of the results from \citet{vandyk+19} and \citet{eldridge+19a}. We do not consider the conclusions of \citet{kilpatrick+18} and \citet{rui+19} with respect to SN~2017eaw as both studies had assumed a shorter distance to the host galaxy (see Sect.~\ref{sec:17eaw}). Finally, we use the mass estimation of \citet{oneill+19} for SN~2018aoq.
In cases where more than one estimation of \mzams\ is available, we take the complete range of values predicted in the different works instead of using some specific value. It should be noted that in almost all of these cases, the most accurate values are within the error bars of other solutions.

Figure \ref{fig:hydro_preexp} compares our results with those mentioned above. We find a good agreement between the masses estimated by both methods for almost every SN in the sample. SN~2004et and SN~2012ec are the only ones that escape from the trend. Our analysis suggests a low-mass progenitor for SN~2012ec while pre-SN imaging propose a more massive one, and the opposite for SN~2004et.

Similar works to that presented here have been published in \citet{morozova+18} and \citet{eldridge+19b}. They use large grids of hydrodynamic simulations and a non-Bayesian fitting procedure based on $\chi^{2}$ minimisation.
\citet{morozova+18} use a large sample of SNe~II which includes eight objects with observed progenitors. At first glance, there is a clear discrepancy between both quantities. They attribute it to the fact that they use pre-SN simulations from a different stellar evolution code to that used to connect the luminosity of the progenitor with its initial mass. \citet{eldridge+19b} only consider the SNe~II with observed progenitors of \citet{smartt15}. They claim that their results are consistent with those from pre-explosion imaging, although they remark that their results have a tendency towards higher masses. 
In summary, both above papers present results that are consistent with those from pre-SN imaging but with a tendency to higher progenitor masses.
It is important to mention that while both works use a larger sample of SNe~II than we do here, some progenitor candidates have not been confirmed yet. Post-explosion images when the SN has faded sufficiently are needed in order to confirm the progenitor through its disappearance. Until this happens, results should be taken with caution as it can lead to a wrong determination of the progenitor star.
Additionally, the mass discrepancy may be due to the fact that the authors only use the LC to obtain the progenitor properties without using any spectral information such as the expansion velocity.
One of the differences with our work is that we fit the photospheric velocity simultaneously with the LC (see Sect.~\ref{sec:degeneracy} for discussion).

In our work, the root-mean-square (RMS) deviation of our progenitor mass estimates against those from direct detections is found to be 2.8~\ms. We now compare with the RMS values from \citet{morozova+18} and \citet{eldridge+19b} to quantify the differences. It must be stressed that when comparing with other works, we recalculate the RMS of our results taking into account only those SNe within both works.
We find a RMS of 3.5~\ms\ in our results and 5.8~\ms\ in those by \citet{morozova+18}. For the results in \citet{eldridge+19b} we estimate a RMS value of 3.4~\ms, while 3.1~\ms\ is found in our work for the same objects. In conclusion, we find that our mass estimations are more consistent with those from pre-explosion imaging than in previous works. 

During the last two decades, tension has emerged between hydrodynamic modelling and pre-SN imaging in the sense that the progenitor mass estimated by the former was usually larger than the estimated or upper limits given by the direct analysis of progenitors. However, we do not find such discrepancy in our analysis.

\subsection{Comparison with results from late-time spectral modelling}
\label{sec:nebular}

\begin{figure}
\centering
\includegraphics[width=0.5\textwidth]{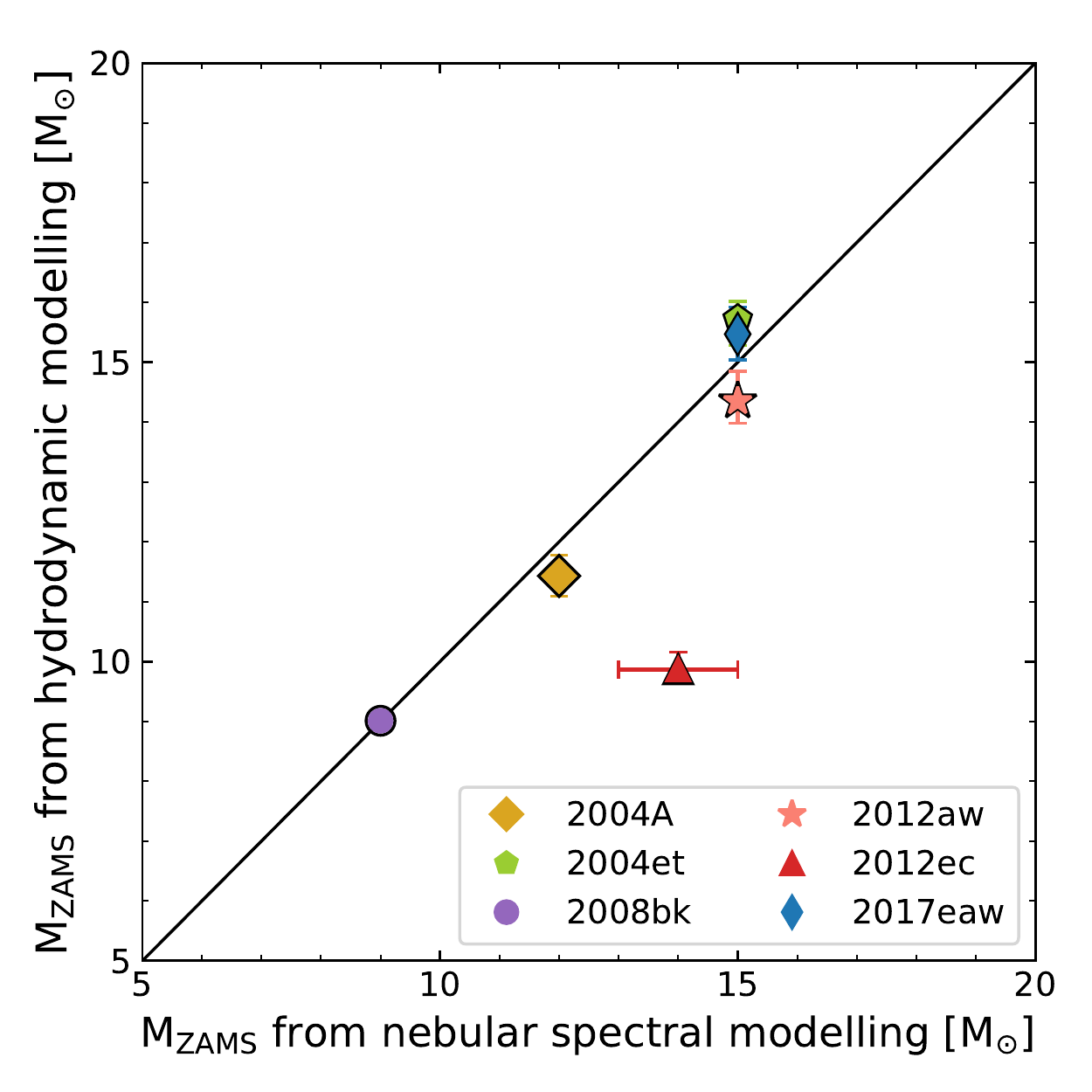}
\caption{Comparison between \mzams\ obtained from hydrodynamic and late-time spectral modelling. An excellent agreement can be seen for all SNe~II except SN~2012ec.}
\label{fig:hydro_nebular}
\end{figure}

We have already compared our results with those which come from the detection and analysis of the progenitor star in pre-explosion images. Similar analysis can be performed by comparing with the results of  progenitor \mzams\ determined through nebular spectral modelling. Late-time spectra of SNe allow examination of the nucleosynthesis yields, especially the emission lines of \ion{[O}{i]} $\lambda\lambda$6300,~6364 as these lines characterise the core mass of the progenitor. Then, by comparison with synthetic nebular spectra available for different \mzams, it is possible to distinguish between different progenitors \citep{jerkstrand+14}.

We use the results from \citet{jerkstrand+12,jerkstrand+14,jerkstrand+15,jerkstrand+18} for SNe~2004et, 2012aw, 2012ec, and 2008bk respectively, \citet{silverman+17} for SN~2004A, and \citet{vandyk+19} for SN~2017eaw to contrast with our values. A comparison of the progenitor masses obtained with both methods are shown in Fig.~\ref{fig:hydro_nebular}. Excellent agreement is found between the two methods for all objects with the exception of SN 2012ec. This is the only one that displays a different solution (see Sect.~\ref{sec:2012ec}).

\subsection{Ejecta masses}

\begin{figure}
\centering
\includegraphics[width=0.5\textwidth]{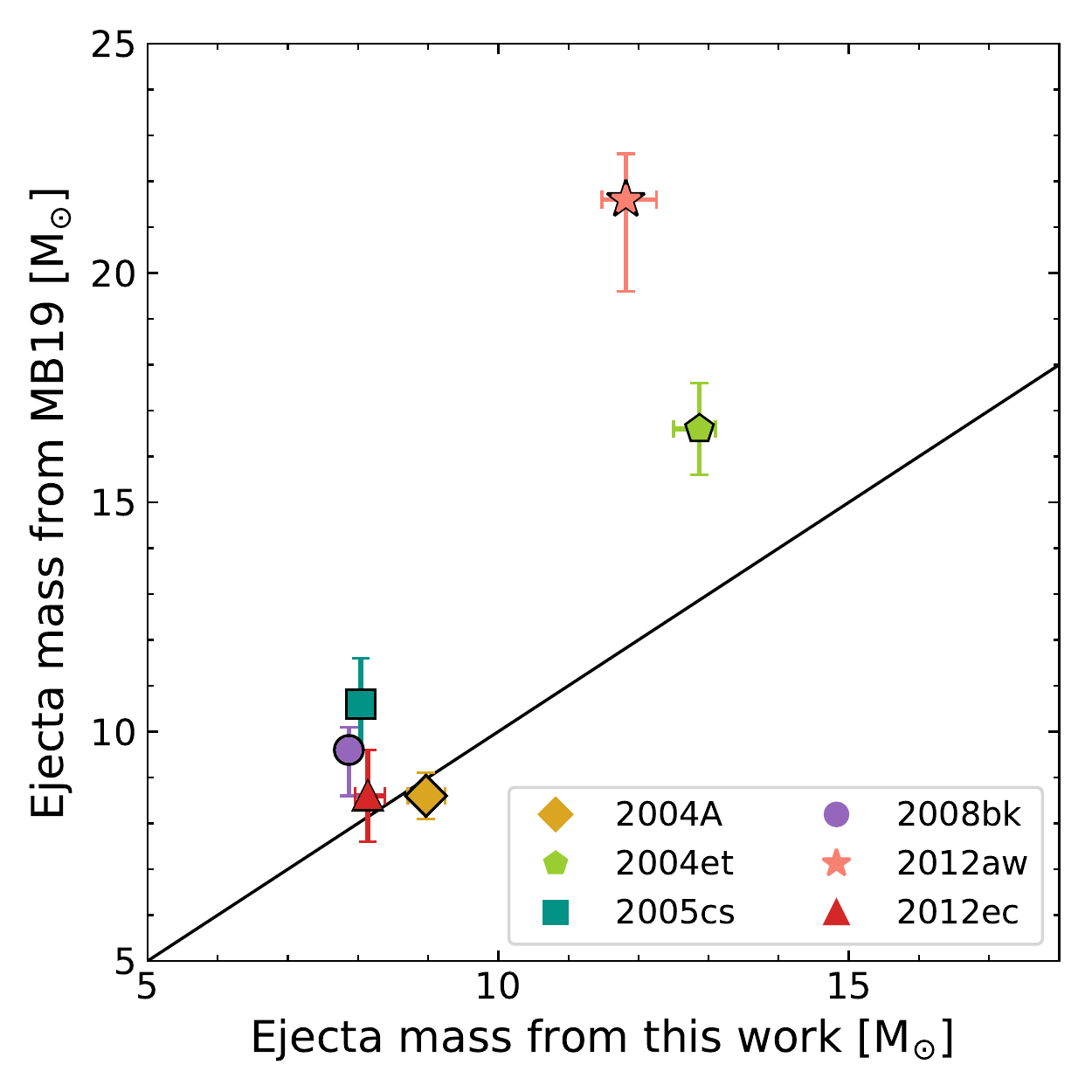}
\caption{Ejecta masses estimated in this work in comparison with those from MB19.}
\label{fig:hydro_ejecta}
\end{figure}

In this section we compare the ejecta masses of this paper with those presented by MB19. Even though both works use the same hydrodynamic code, the progenitor models and the selection of the preferred model was done in a very different way. In MB19 double polytropic models were used as pre-SN structures, while here we use stellar evolution calculations.
Polytropic models allow to obtain the final structure of the star dependent only on the pre-SN properties, a priori without information of the initial mass. Therefore, we only provide a comparison to the ejecta masses instead of main-sequence masses. In addition, MB19 chose their preferred models by visual comparison while here we use a robust statistical method.

Figure~\ref{fig:hydro_ejecta} shows a poor agreement between both ejecta masses. In fact, the differences are systematic in the sense that the ejecta masses in MB19 are larger than ours. SN~2004et and SN~2012aw present the largest differences being of the order of 3.5~\ms\ and 10~\ms, respectively. As we mentioned above, the only physical difference between both works is in the calculation of the pre-SN models. Double polytropic calculations allow to produce a large variety of pre-SN structures with different mass, radius, chemical composition, and density profiles.
The pre-SN models in MB19 may indicate that different solutions could be found if the standard assumptions in stellar evolution change, for example with respect to mixing processes and mass-loss rates.
This is an option given the uncertainties still present in stellar modelling, especially in massive stars.
Additionally, in this work we use a large grid of simulations and a fitting method with statistical support that initially covers the entire parameter space. While visual comparison can find solutions that reproduce the observations, it does not consider whether other solutions are possible - sometimes even more probable solutions.

\subsection{Progenitor radii}

\begin{figure}
\centering
\includegraphics[width=0.5\textwidth]{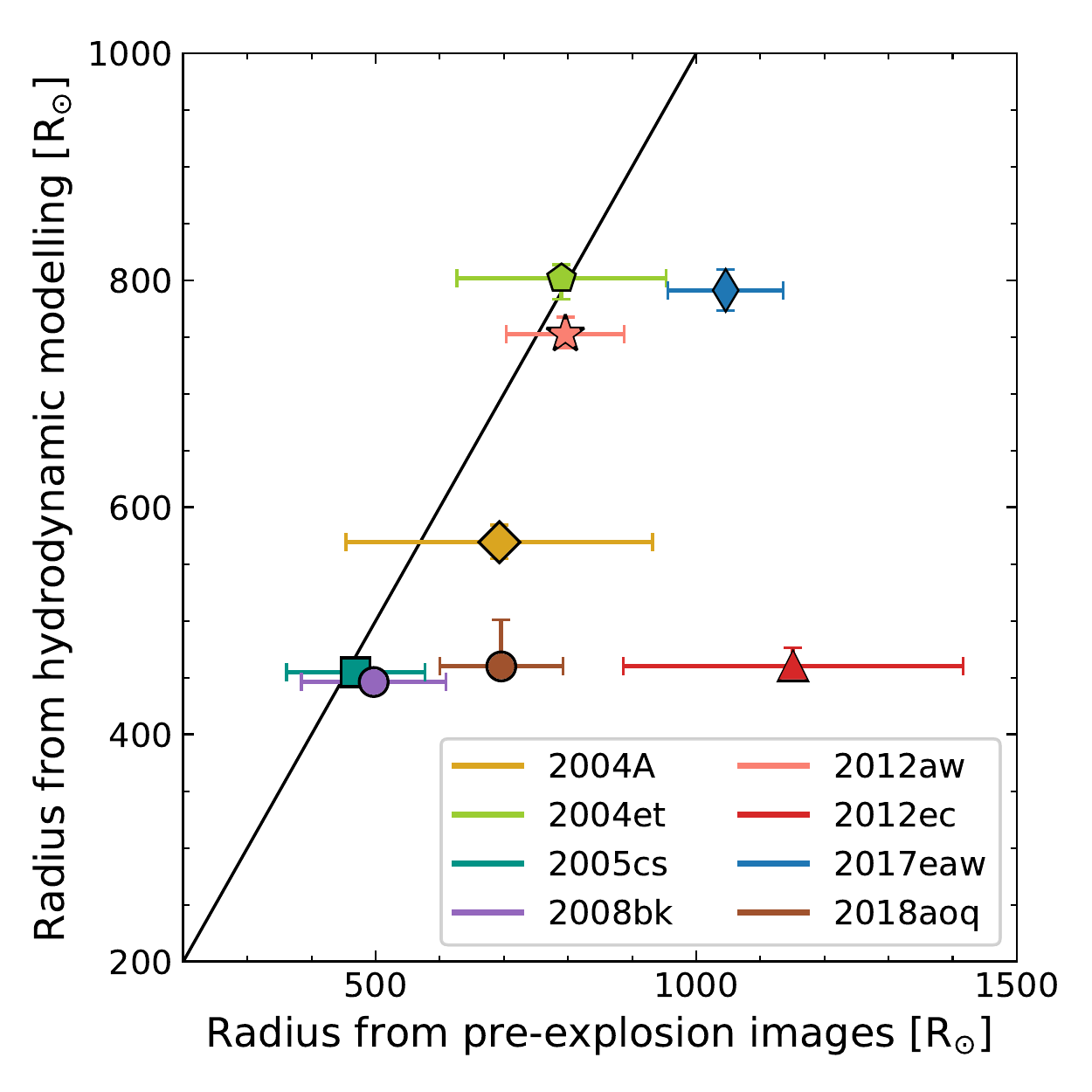}
\caption{Comparison between progenitor radii from our modelling with that derived from observed progenitor properties before explosion.}
\label{fig:hydro_radius}
\end{figure}

Direct detections provide an unique opportunity to place the progenitor in an HR diagram. The observations determine the progenitor luminosity and effective temperature. Additionally, assuming a black body, the progenitor radius can be estimated. On the other hand, we can recover the final radius, as well as other properties, from the \mzams\ derived from the hydrodynamical modelling as we use progenitor models from stellar evolution calculations.

In the following we compare the progenitor radius from pre-SN imaging to our estimations. We use the values of the luminosity and effective temperature from \citet{smartt15} to estimate the progenitor radius for SN~2004A, SN~2004et, SN~2005cs, SN~2008bk, SN~2012aw, and SN~2012ec, and \citet{oneill+19} for SN~2018aoq. Finally, we use the radius of the progenitor of SN~2017eaw from \citet{vandyk+19}. Progenitor radii for each pre-SN model is listed in Table~\ref{table:presn_models}. The progenitor radius was interpolated linearly to the \mzams\ we derived from the fitting.

Figure~\ref{fig:hydro_radius} shows the comparison between progenitor radii from our modelling with that derived from observed progenitor properties before explosion. We note that our results are consistent for five objects in our sample, while our modelling underpredicts the remaining three radii. In general, there is a tendency to yield lower progenitor final radius by an average of 175~\rs.
Nevertheless, the agreement is quite good considering that the progenitor radius is not a free parameter in our modelling.

\section{Discussion}
\label{sec:discussion}

\begin{figure*}
\centering
\includegraphics[width=0.85\textwidth]{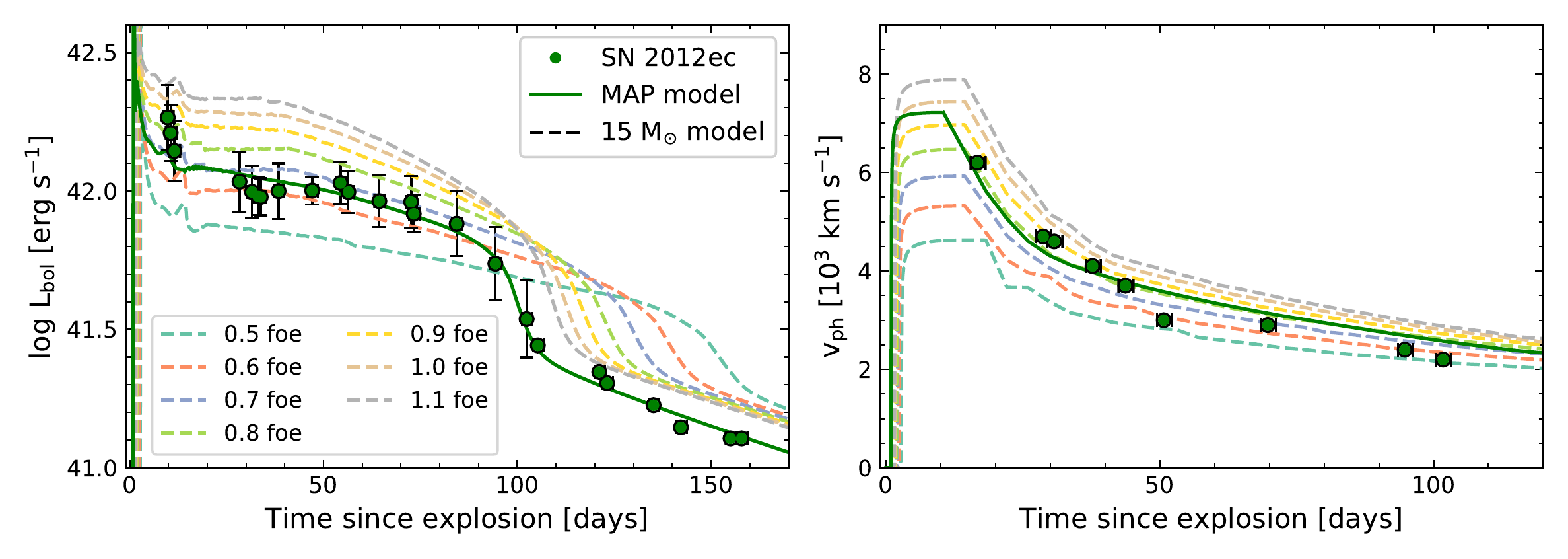}
\includegraphics[width=0.85\textwidth]{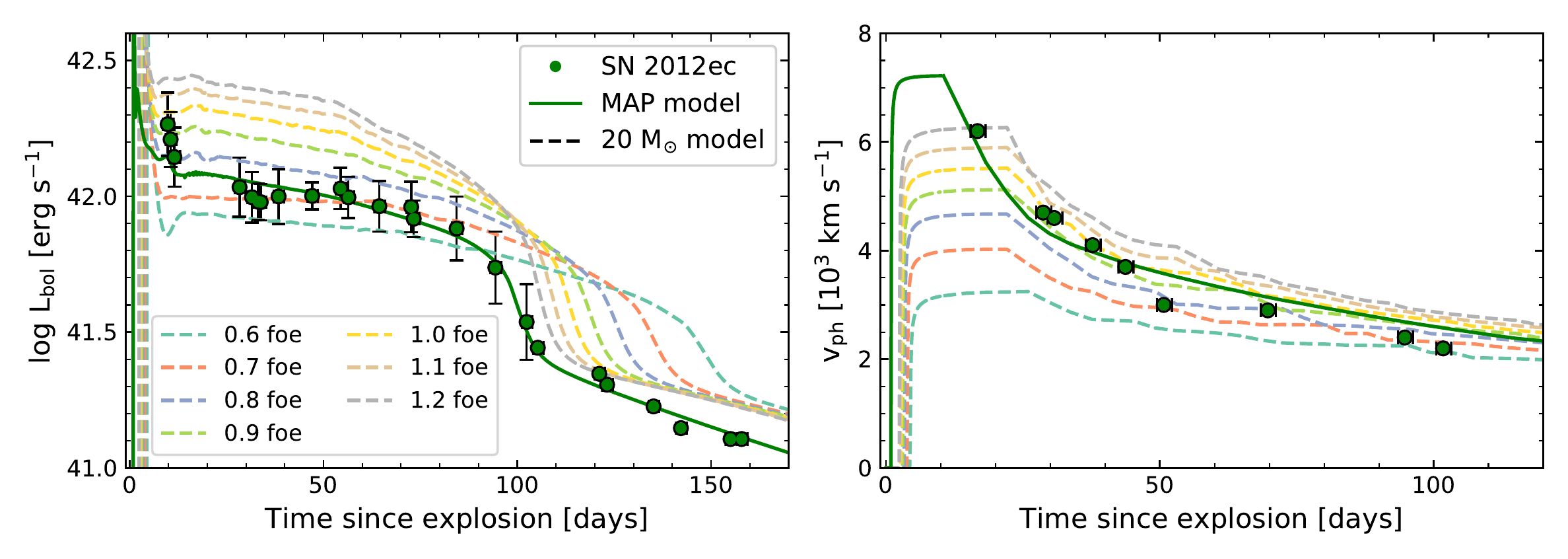}
\includegraphics[width=0.85\textwidth]{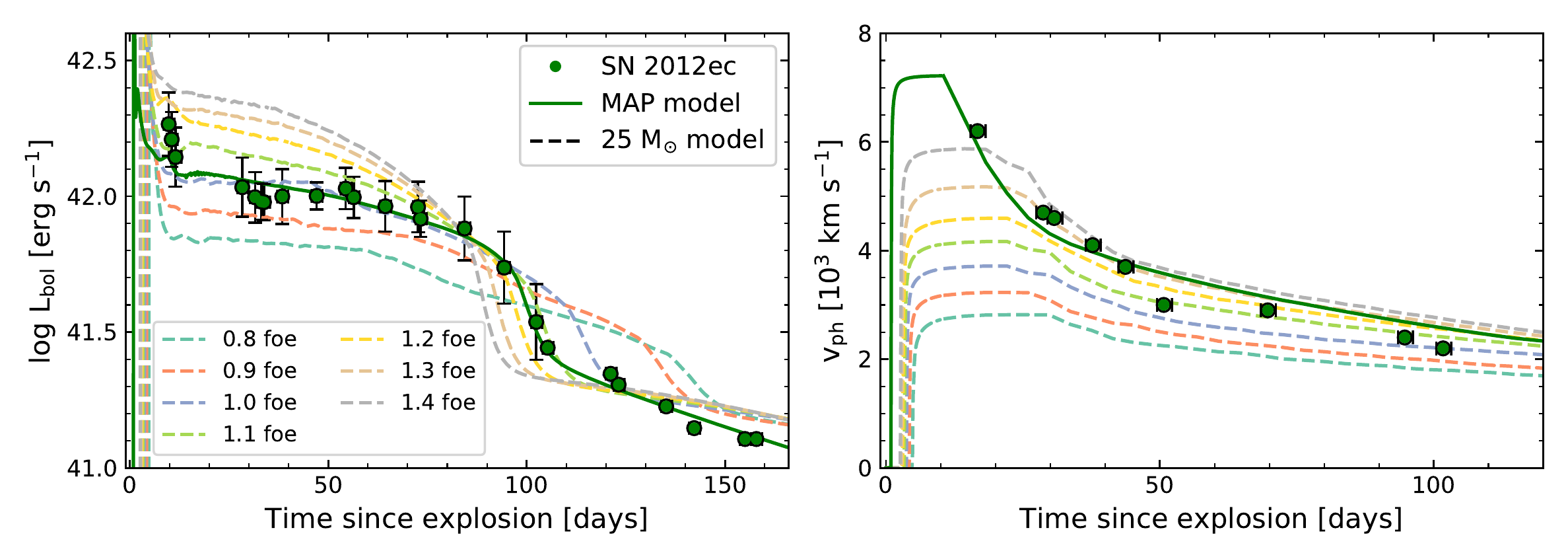}
\caption{Hydrodynamic models for different \mzams\ (dashed lines) compared to the MAP model (solid line) and observations of SN~2012ec (green dots). Dashed lines are colour-coded according to different explosion energy. The models plotted correspond to the same values of \mni\ and \mix\ as the MAP model. The MAP model is reproduced with \mzams~=~9.9~\ms, $E$~=~0.57~foe, \mni~=~0.034~\ms, and \mix~=~28\%. \textit{Left panels:} bolometric LCs. \textit{Right panels:} evolution of the photospheric velocity. \textit{Top panel:} 15~\ms\ model. \textit{Middle panel:} 20~\ms\ model. \textit{Bottom panel:} 25~\ms\ model.}
\label{fig:high_mass}
\end{figure*}

\begin{figure*}
\centering
\includegraphics[width=1.0\textwidth]{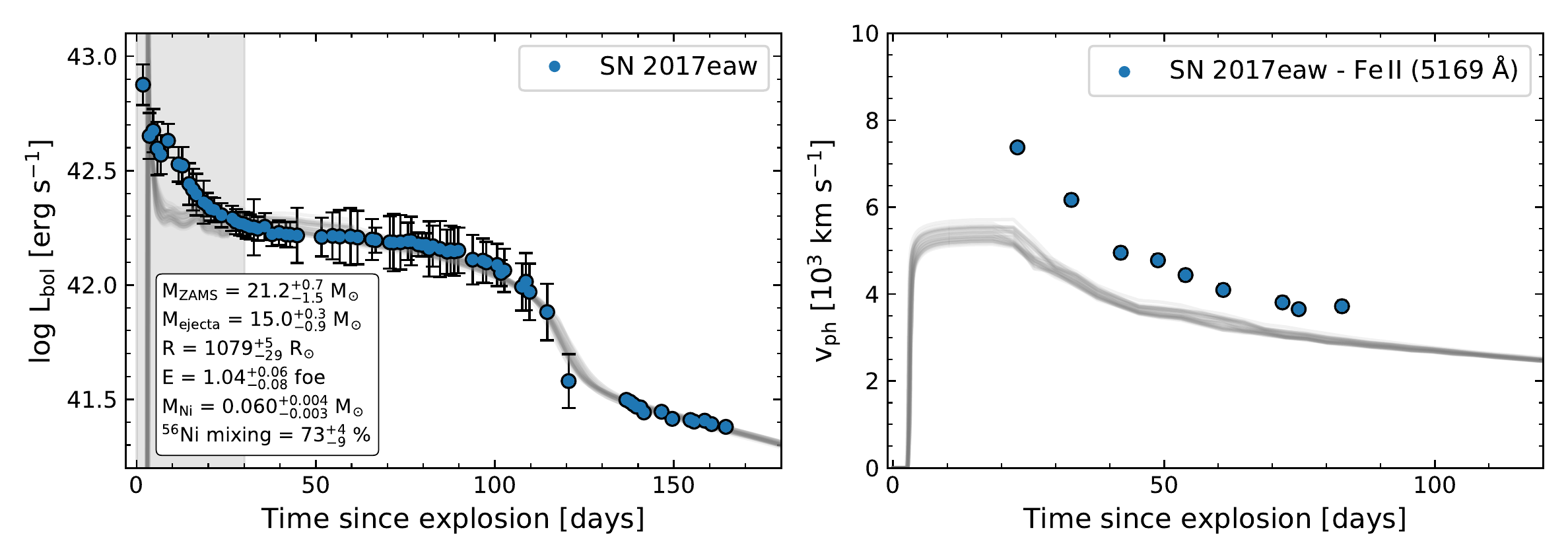}
\caption{Bolometric LC (left panel) and photospheric velocity evolution (right panel) for SN~2017eaw in comparison with fifty hydrodynamical models randomly chosen from the posterior probability distribution. In this example we did not take into account the photospheric velocity for the fitting. We can note the importance of fitting both observables simultaneously.}
\label{fig:17eaw_no_vel}
\end{figure*}

\subsection{LC degeneracies}
\label{sec:degeneracy}

One of the main problems of inferring progenitor properties from LC modelling is that there is a degeneracy among some progenitor properties when reproducing the observations. Sometimes, similar photometric properties can be achieved with different progenitor and explosion characteristics. 
In \citet{dessart+19}, the authors pointed out that different progenitors can finish with a comparable H-rich envelope mass and produce similar photospheric phases. Moreover, \citet{goldberg+19} argue that the ejecta mass, the explosion energy, and the progenitor radius cannot be constrained from the LC and velocity measurements. They solve it using two of these explosion properties as a function of the third. This requires independent knowledge of one parameter.

During the last years, hydrodynamic modelling of SNe~II LCs and velocity measurements has suggested a significant discrepancy between the SN ejected masses and the initial masses of the observed progenitors.
For example, \citet{utrobin+08} present detailed modelling of SN~2005cs and a summary for other three SNe~II (1987A, 1999em, and 2003Z). They argue that the hydrodynamic progenitor masses are systematically higher than if SNe~II had originated from the range of 9--25~\ms, assuming a Salpeter initial mass function. This differs markedly with the direct detection of progenitors in pre-explosion images.
However, MB19 analysed a sample of six SNe~II with confirmed progenitors in post-explosion images and find that hydrodynamic masses are not systematically larger that those from pre-SN imaging. But they do note that using similar pre-SN models and explosion parameters as \citet{utrobin+08} and \citet{utrobin+09}, they arrive at similar LCs. Once again, this shows the high degree of degeneracy present in this problem.
We consider that a detailed inspection of the degeneracy and the discrepancy between hydrodynamic masses and the masses inferred by the direct detection of the progenitors can be achieved with a large grid of hydrodynamic models in parameter space consisting of a considerable variety of LC and velocity models and a robust fitting procedure with statistical support, as presented in this work.
With the above-mentioned considerations and from the analysis presented in Sect.~\ref{sec:preexp} we conclude that we do not find such discrepancy between the progenitor initial masses inferred by hydrodynamical modelling and pre-SN imaging.

Furthermore, we also compare high-mass models with the observed LCs and photospheric velocities to examine how different they look.
We show the case of SN~2012ec as example, but we performed the same analysis for the complete sample.
Figure~\ref{fig:high_mass} shows hydrodynamic models for different initial masses (dashed lines) compared to the MAP model for SN~2012ec (solid line). In addition, models for several explosion energies were plotted. Every model has the same \mni\ and \mix\ as the MAP model.
The top panel of Fig.~\ref{fig:high_mass} shows the case for a 15~\ms\ progenitor. Despite the large range of explosion energies plotted we note that none of the models reproduce the LC. We do not show a larger range of energies since more energetic models will produce brighter and shorter plateau phases, and the opposite for models with lower explosion energy. The plateau luminosity and the photospheric velocity are well represented by the model with an explosion energy of 0.7~foe, but it predicts a longer plateau duration ($\sim$40~days longer). 
Something similar is seen in the middle panel of Fig.~\ref{fig:high_mass} for the 20~\ms\ model. The main difference is that in this example the model with 0.7~foe of energy also fails to reproduce the expansion velocities at early times.
The bottom panel shows models computed for a 25~\ms\ progenitor. Here, the model with 1.1~foe reproduces the plateau length but it fails in the plateau shape. Additionally, this model predicts low expansion velocities which differ from the observations.
We conclude that we are not missing high-mass solutions. Therefore, we feel confident that our fitting procedure finds the best solutions and, at least within the grid of models we are using, other models are much less probable.

In the case of the 25~\ms\ progenitor, the contribution of the observed velocities when discarding the models with explosion energies between 1.0 and 1.1~foe is considerable.
In the following we show with a simple test how, in some cases, the photospheric velocity helps to solve the dichotomy between different solutions. For this purpose we ran the MCMC sampler for SN~2017eaw again but this time considering only the bolometric LC. Results can be seen in Fig.~\ref{fig:17eaw_no_vel}. As expected, we obtain a good agreement to the LC while photospheric velocities are not well reproduced. In this case the estimated \mzams\ is 21.2~\ms, that is $\sim$6~\ms\ larger than the estimated through LC and photosperic velocity modelling.

Clearly, fits to the LC alone is not a good method to estimate the physical properties of explosions. At least in some cases the photospheric velocity evolution is essential in breaking the degeneracy \citep[see also][]{ricks+19}.
However, this is the methodology used recently by \citet{morozova+18} and \citet{eldridge+19b}. In both papers, there is a tendency to higher progenitor masses with respect to the stellar masses based on direct analysis of progenitors in pre-explosion imaging. 
With this example we thus emphasise that care has to be taken when deriving progenitor properties without using any spectral information. If this observable is not taken into account, it could lead to a wrong determination of the mass and energy.

\subsection{Limitations and caveats}
\label{sec:limitations}

\begin{figure}
\centering
\includegraphics[width=0.5\textwidth]{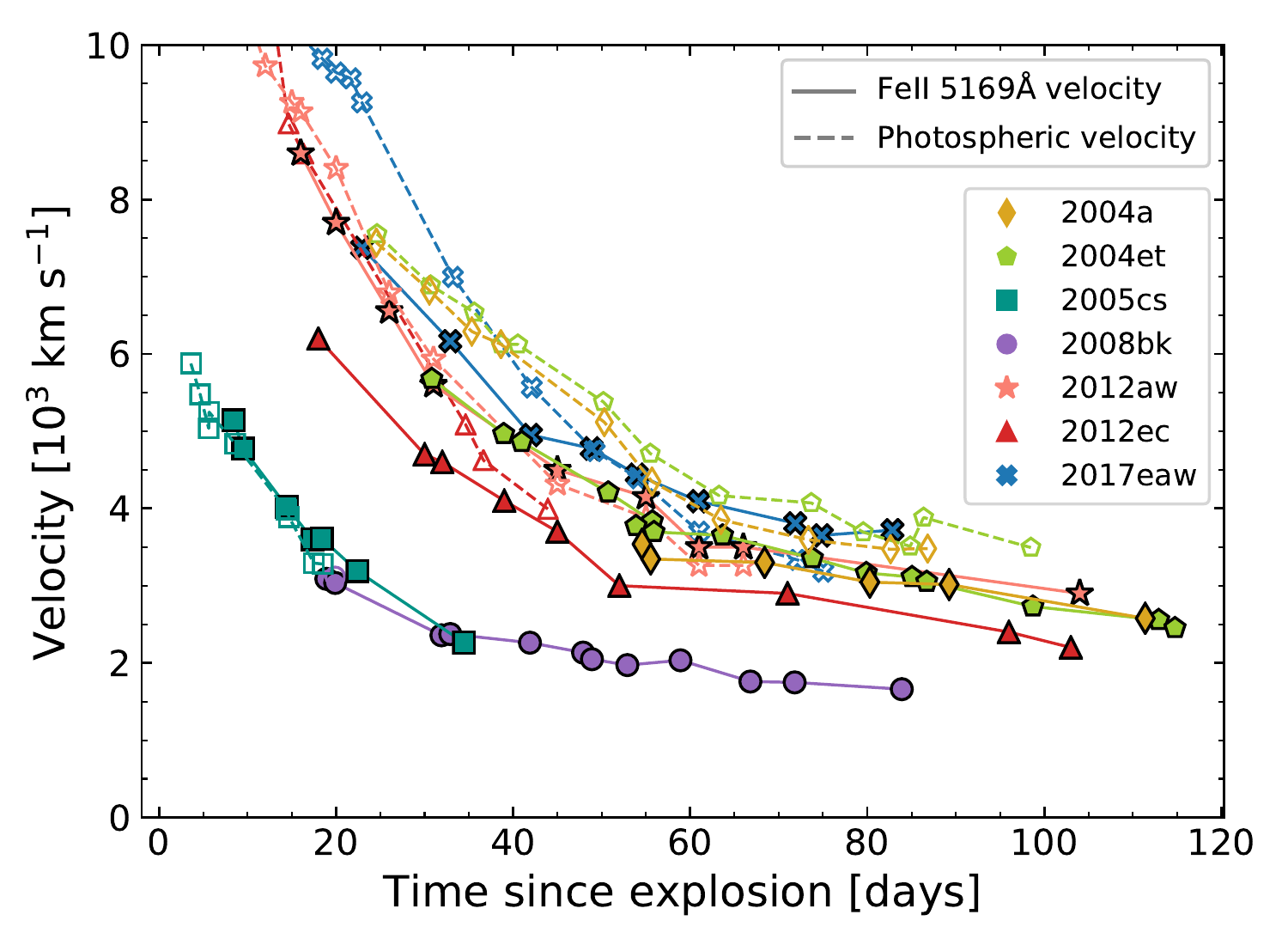}
\caption{Comparison between observed \ion{Fe}{ii} velocities (filled markers, solid lines) and photospheric velocities (open markers, dashed lines) inferred from H$\beta$ velocities through the polynomial relation from \citet{jones+09} for each SN in our sample. Errors are not plotted for better visualisation.}
\label{fig:obsfeii_vph}
\end{figure}

\begin{figure}
\centering
\includegraphics[width=0.5\textwidth]{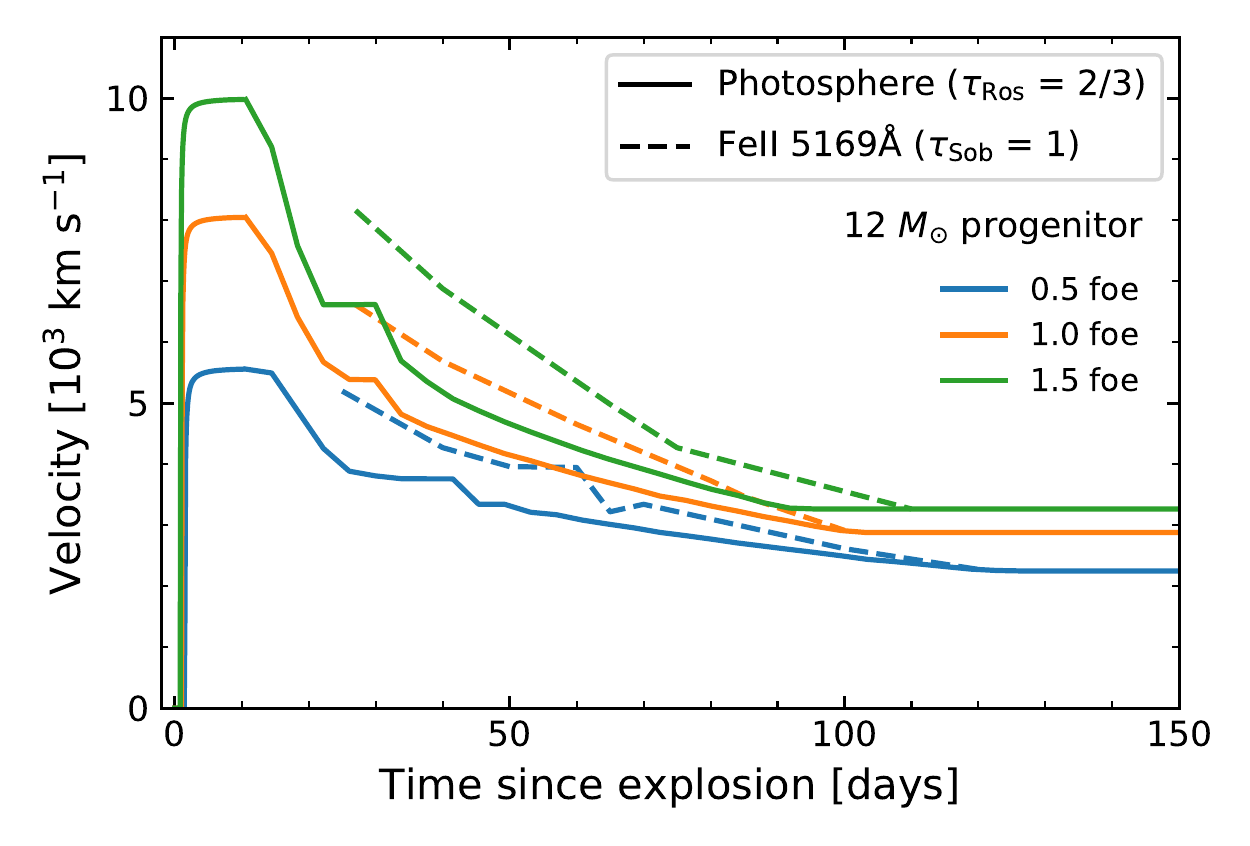}
\caption{Comparison of the photospheric velocity and the \ion{Fe}{ii} line velocity for a 12~\ms\ progenitor and different explosion energies.}
\label{fig:feii_velocities}
\end{figure}

The stellar evolution simulations presented in this work require a large number of assumptions. In the calculation of our progenitor models we assume non-rotating stars and standard values for the mixing-length parameter and overshooting. \citet{dessart+13} explore how variations of these parameters affect  the final structure of a SN~II progenitor. Different values produce changes in the progenitor radius, the H-rich envelope mass, and the helium-core mass, among others, which influence significantly the LCs, although the initial mass is the same.
Studying all the existing possibilities, according to the different evolutionary parameters that can be used, is difficult and beyond the scope of this work.

As we have mentioned in Sect.~\ref{sec:models}, the hydrodynamic code we use assumes a radially symmetric flow and adopts LTE to describe the radiative transfer. Although these approximations might not be entirely correct, they seem to be a good approach. The very extended and massive hydrogen envelopes that characterise SNe~II are expected to smooth the asymmetries of the explosion mechanism which makes spherical symmetry a good approximation for the bulk of the ejecta, though the \Ni\ distribution is more likely to be in some preferred direction \citep{wongwathanarat+15}. On the other hand, LTE assumes that radiation and matter are strongly coupled. This is not valid at shock breakout and during and after the transition phase to a completely recombined ejecta.

Another approximation in the code is regarding the opacity calculation. The code uses opacity tables calculated assuming LTE and a medium at rest. These calculations underestimate the true line opacity when considering rapidly expanding envelopes where large velocity gradients are present \citep{karp+77}. In addition, the effect of the non-thermal excitation or ionisation of electrons that are created by Compton scattering of $\gamma$-rays emitted by radioactive decay of \Ni\, and $^{56}$Co is not included in the calculation of the opacity. Our assumptions in the calculation of the opacity considerably underestimates the true ionisation. To partially solve the underestimation in the opacity, the code adopts a minimum value of the opacity sometimes referred as the ``opacity floor'' \citep[see more details in][]{bersten+11}. This approach has been extensively used in the literature \citep[see, e.g.][]{young04,morozova+15}. 

The analytical study of \citet{popov+93} shows the dependence of the bolometric luminosity on the opacity as it also shows the subsequent dependence of the explosion energy, mass and radius on the opacity.
Qualitatively, a larger value of the opacity decreases the plateau luminosity while increasing the duration of the plateau. As a consequence, this leads to different progenitor and explosion parameters.
From Eq.~27 of \citet{popov+93}, if the bolometric luminosity is fixed, an increase in the opacity leads to a higher explosion energy and a lower mass.
Thus, given the opacity has an important effect on the calculation of the bolometric LC models, this gives rise to a corresponding uncertainty in the progenitor and explosion parameters.\\

Finally in this section we discuss the uncertainties in our results that arise from the assumed photospheric velocities from both the models and the observations.
As mentioned in Sect.~\ref{sec:sample}, our hydrodynamical modelling requires the measurement of the ejecta photospheric velocity. One of the typical procedures to estimate the photospheric velocity is by measuring the velocity at maximum absorption of optically-thin lines as it is assumed that these lines are formed near the photosphere \citep{leonard+02}. 
\citet{dessart+05} analysed several synthetic line velocities and determined that the \ion{Fe}{ii}~5169~\AA\ line delivers high accuracy in matching the photospheric velocity.
This assumption is extensively used in the literature. Consequently, we use this line velocity as photospheric velocity indicator.  However, the results achieved by \citet{dessart+05} are restricted to a minimum velocity of $\sim$4000~km~s$^{-1}$.
Since some of the \ion{Fe}{ii} velocities in our sample are below that limit, we now discuss and analyse how the use of different techniques for estimating this velocity may affect our results.
\citet{jones+09} used synthetic spectra from \citet{eastman+96} and \citet{dessart+05}, and found polynomial relations to convert the observed H$_{\beta}$ velocities into photospheric velocities. Moreover, they suggest that models from \citet{eastman+96} predict more realistic line profiles in the SN ejecta than \citet{dessart+05}, and therefore should provide a better photospheric velocity estimation. For this reason, we use the relation found by \citet{jones+09} with models from \citet{eastman+96} to derive the photospheric velocity. 
Figure~\ref{fig:obsfeii_vph} compares this photospheric velocity with the observed \ion{Fe}{ii} velocities.
Within the uncertainties both velocities are generally consistent for the full sample. 
There are, however, some caveats: a) despite the good agreement for SN~2008bk, only one photospheric velocity is available for comparison as only one measure of the H$_{\beta}$ velocity is within the range of validity of the polynomial relation, b) H$_{\beta}$ velocities are not available for SN~2018aoq, and c) for SN~2004et, there are differences between \ion{Fe}{ii} and photospheric velocities of the order of 1000~km~s$^{-1}$ at 30$-$40~days after explosion. These differences decrease with time.
This comparison provides additional support to the use of \ion{Fe}{ii} velocities as photospheric velocities. 
However, an additionally caveat is that
the photospheric position in our models can be different from those in atmospheric models. This is due to the differences in the opacities involved in determining this location. In our models, the photosphere is defined where the Rosseland mean optical-depth is 2/3, while in \citet{eastman+96} the photosphere is located where the Thomson scattering optical-depth equals 2/3.

Recent studies have used another approximation in order to model ejecta velocities. Instead of using observed line velocities as photospheric velocity indicators, these works calculate the \ion{Fe}{ii} line velocity in the Sobolev approximation where the Sobolev optical depth equals one \citep{paxton+18,goldberg+19,ricks+19,bostroem+19}. However, the precise Sobolev optical depth where the line is formed is not known, and different values translate into different velocities \citep[see Fig.~36 of][]{paxton+18}.
We compare our model photospheric velocity with the \ion{Fe}{ii} line velocity as defined by these authors to check how our results could be influenced by this issue. We use Eq.~53 of \citet{paxton+18} to calculate the Sobolev optical depth for the \ion{Fe}{ii} line. In order to do so, the ionisation fraction of iron atoms is needed. This information is provided in a table as a function of density and temperature and was obtained from the public version of \texttt{MESA}. The expression for the Sobolev optical-depth \citep[Eq.~53 of][]{paxton+18} is valid in a homologously expanding atmosphere. Therefore, the \ion{Fe}{ii} velocities are calculated only for times later than 25~days after explosion.

Figure~\ref{fig:feii_velocities} compares models of photospheric velocity and \ion{Fe}{ii} velocity for a 12~\ms\ progenitor and three values of the explosion energy: 0.5, 1.0, and 1.5~foe. Differences of about 1000~km~s$^{-1}$ are found at early times.
As time goes by, the differences decrease.
Therefore, if this way of comparing model \ion{Fe}{ii} velocities with observations is precise, we expect some changes in our results.
According to Fig.~\ref{fig:feii_velocities}, our photospheric velocity models underestimate \ion{Fe}{ii} velocities.
As the expansion velocities are mostly affected by the energy of the explosion, our results could overestimate the explosion energy. Changes in progenitor masses are also possible.

The following analysis estimates how our results could change if observed \ion{Fe}{ii} velocities are fitted using these model \ion{Fe}{ii} velocities.
We compute bolometric LC and \ion{Fe}{ii} velocity models for 12, 15, 18, and 20~\ms\ progenitors, with explosion energies of 0.5, 0.8, 1.0, and 1.2~foe for each mass value.
Then, we fit these LCs and \ion{Fe}{ii} velocity models using the grid of models described in Sect.~\ref{sec:models}, i.e. using the photospheric velocity models. This analysis gives an estimation of the differences we can expect in our results if \ion{Fe}{ii} velocity models are used to compare with the observations. 
We find a tendency to yield larger ejecta masses and explosion energies by an average of $\sim$0.2~\ms\ and $\sim$0.2~foe, respectively. 
Changes in our results would go in the opposite direction as photospheric velocities are slower than \ion{Fe}{ii} line velocities (Fig.~\ref{fig:feii_velocities}).
Therefore, how one defines model velocities for comparison to observations can be associated to a systematic error in the ejecta masses and explosion energies.
However, while our results on individual SNe~II would change moderately, this would not significantly alter our conclusions.

In conclusion to this analysis, we have discussed different procedures to estimate the ejecta photospheric velocity, as well as other techniques to model the ejecta velocities. While small differences in best-fit physical parameters emerge, it is not completely clear which model velocities one should use, as both have uncertainties. The differences found can introduce a possible small bias in our results, but it does not affect our conclusions.

\subsection{SN~2012ec}
\label{sec:2012ec}

In Sects.~\ref{sec:preexp} and \ref{sec:nebular} we compare the progenitor masses we obtained from the hydrodynamic modelling with the estimations from the direct detection of the progenitor star in pre-explosion images and nebular spectral modelling recovering a strong agreement. However, SN~2012ec is the only one that shows different solutions in both cases. From the hydrodynamic modelling point of view, we conclude an ejecta mass of $\sim$8~\ms, related to a main-sequence star of $\sim$10~\ms. On the other hand, pre-SN imaging infers a luminous and red progenitor. It is worth emphasising that pre-SN imaging provide a luminosity range and not the mass. The luminosity is then converted to an initial mass after comparison with evolutionary tracks of single-star models of different initial masses that terminates in the RSG phase within that range in luminosity. This analysis shows a progenitor of 16~$\pm$~5~\ms.
We observe that despite being more massive than our value, both estimates are not statistically distinct, given the large uncertainty of the pre-explosion image analysis.
Furthermore, nebular spectral modelling suggests a main-sequence mass range of 13--15~\ms\ due to the core mass of the progenitor. 
If we collect all this information we obtain a luminous progenitor with a core mass corresponding to those stars of $\sim$14~\ms\ in the ZAMS, but with a typical ejecta mass of a single 10~\ms\ star.
A detailed analysis of the progenitor star of SN~2012ec is beyond the scope on this work. 
Nevertheless, we note that the above combination can be obtained if we assume an enhanced rotation or binarity for SN~2012ec, as proposed by \citet{straniero+19}. Rotation produces higher mass helium cores and lower H-rich envelopes. A binary system with a primary star of initial mass estimated by nebular spectral modelling that experiences mass transfer episodes could also explain the disparity. In addition, from analytical estimations and by performing population synthesis simulations, \citet{zapartas+19} conclude that a significant fraction (from $1/3$ to $1/2$) of SN II progenitors are expected to interact with a companion before exploding, which supports this idea.

\section{Summary and conclusions}
\label{sec:conclusions}

We have calculated a large grid of hydrodynamic models applied to stellar evolution progenitors in order to study the nature of SNe~II. LC modelling can provide constraints on progenitor and explosion properties although, sometimes, there is no unique solution. Therefore, we develop a robust method to derive physical properties based on MCMC methods by using the observed bolometric LC and the expansion velocity simultaneously.

We applied this method to the observations of a well-studied set of SNe~II (SNe~2004A, 2004et, 2005cs, 2008bk, 2012aw, and 2012ec) with the aim of comparing with previous results from the analysis of pre-SN imaging. These SNe present confirmation of the progenitor identification via its disappearance in post-explosion images. In addition, we also include SN~2017eaw and SN~2018aoq to the sample, as these are the last SNe to be discovered and analysed with this method. We find that our results are entirely consistent between the initial masses estimated by both methods for almost every SN in the sample. 
Moreover, some works have questioned the ability of the hydrodynamic modelling to recover progenitor and explosion parameters, in the sense that progenitor masses from hydrodynamic modelling are usually larger.  With this analysis we discard such a discrepancy and find a robust method to recover the progenitor mass, among other progenitor and explosion properties.
We note that future high-resolution observations of the explosion site of SN~2017eaw and SN~2018aoq will be required in order to confirm their progenitor candidates.
Additional comparison between our progenitor mass estimations and those from nebular spectral modelling was also computed showing a very good agreement between these methods.

From the proposed analysis we conclude that we have developed a robust method to infer progenitor and explosion properties of SN II progenitors which is in complete agreement with results derived from other methods.
We are now confident in our method and we are able to move on to analyse a larger sample of SNe II.

\begin{acknowledgements}
We thank the anonymous referee for providing constructive comments that improved the content of this paper.
SGG acknowledges support by FCT under Project CRISP PTDC/FIS-AST-31546 and UIDB/00099/2020. LM acknowledges support by UNRN under Project PI2018-40B696.
\end{acknowledgements}

\textit{Software:} \texttt{emcee} \citep{emcee}, \texttt{corner.py} \citep{corner}, \texttt{matplotlib} \citep{matplotlib}, \texttt{MESA} \citep{paxton+11,paxton+13,paxton+15,paxton+18,paxton+19}, \texttt{MesaScript} \citep{mesascript}, \texttt{jupyter} \citep{kluyver2016jupyter}, \texttt{pandas} \citep{pandas}.

\bibliographystyle{aa}
\bibliography{biblio.bib}

\begin{appendix}
\section{Additional plots}

\subsection{Interpolation method}

\begin{figure*}
\centering
\includegraphics[width=0.9\textwidth]{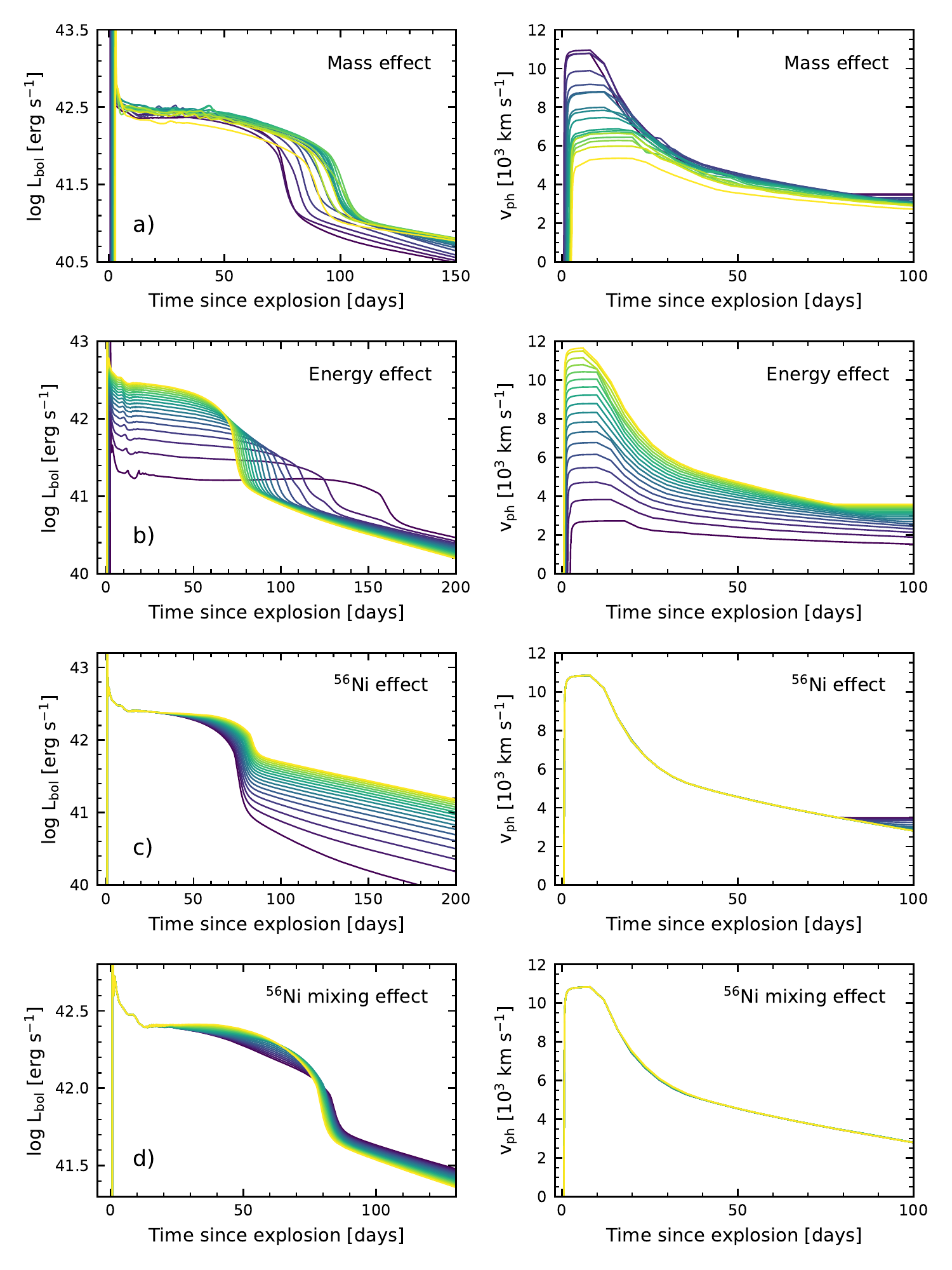}
\caption{Interpolated synthetic bolometric LCs (left panels) and photospheric velocities (right panels) after varying the progenitor initial mass between 9 and 25~\ms\ (panel~a), the explosion energy between 0.1 and 1.5~foe (panel~b), the \Ni\ mass between 0.005 and 0.08~\ms\ (panel~c), and the \mix\ between the 20\% and 80\% of the final structure in mass coordinate (panel~d). The values increase from purple to yellow. Eighteen models are shown for each parameter been varied. The parameters not being varied are fixed at an initial mass of 10~\ms, explosion energy of 1.3~foe, \mni\ of 0.01~\ms, and 50\% of \mix, with the exception of the panel showing the \mix\ effect for which a larger \mni\ of 0.06~\ms\ is used to enable better visualisation of this effect.}
\label{fig:interpolation}
\end{figure*}

In this section, some examples of interpolated hydrodynamical models of bolometric LCs and photospheric velocities are presented.
Fig~\ref{fig:interpolation} shows interpolated models after varying \mzams, \emph{E}, \mni, and \mix. 

\subsection{Corner plots of the posterior probability distributions}
\label{ap:corner}

\begin{figure}
\centering
\includegraphics[width=0.5\textwidth]{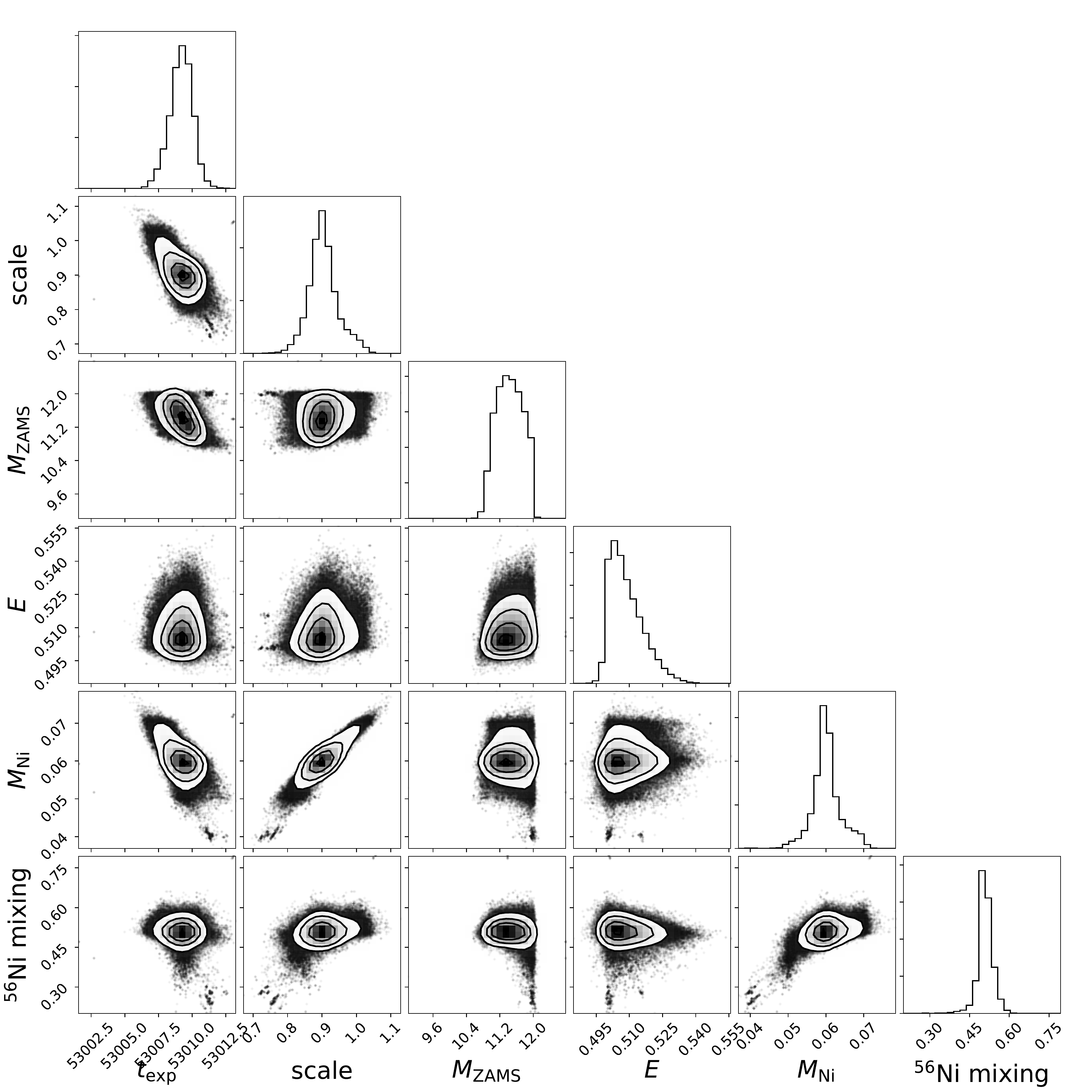}
\caption{Same as for Fig.~\ref{fig:corner_12ec} but for SN~2004A.}
\label{fig:corner_04a}
\end{figure}

Additional corner plots of the joint posterior probability distribution of the parameters are presented in Figs.~\ref{fig:corner_04a} to \ref{fig:corner_12ec_texp_scale}.

\begin{figure}
\centering
\includegraphics[width=0.5\textwidth]{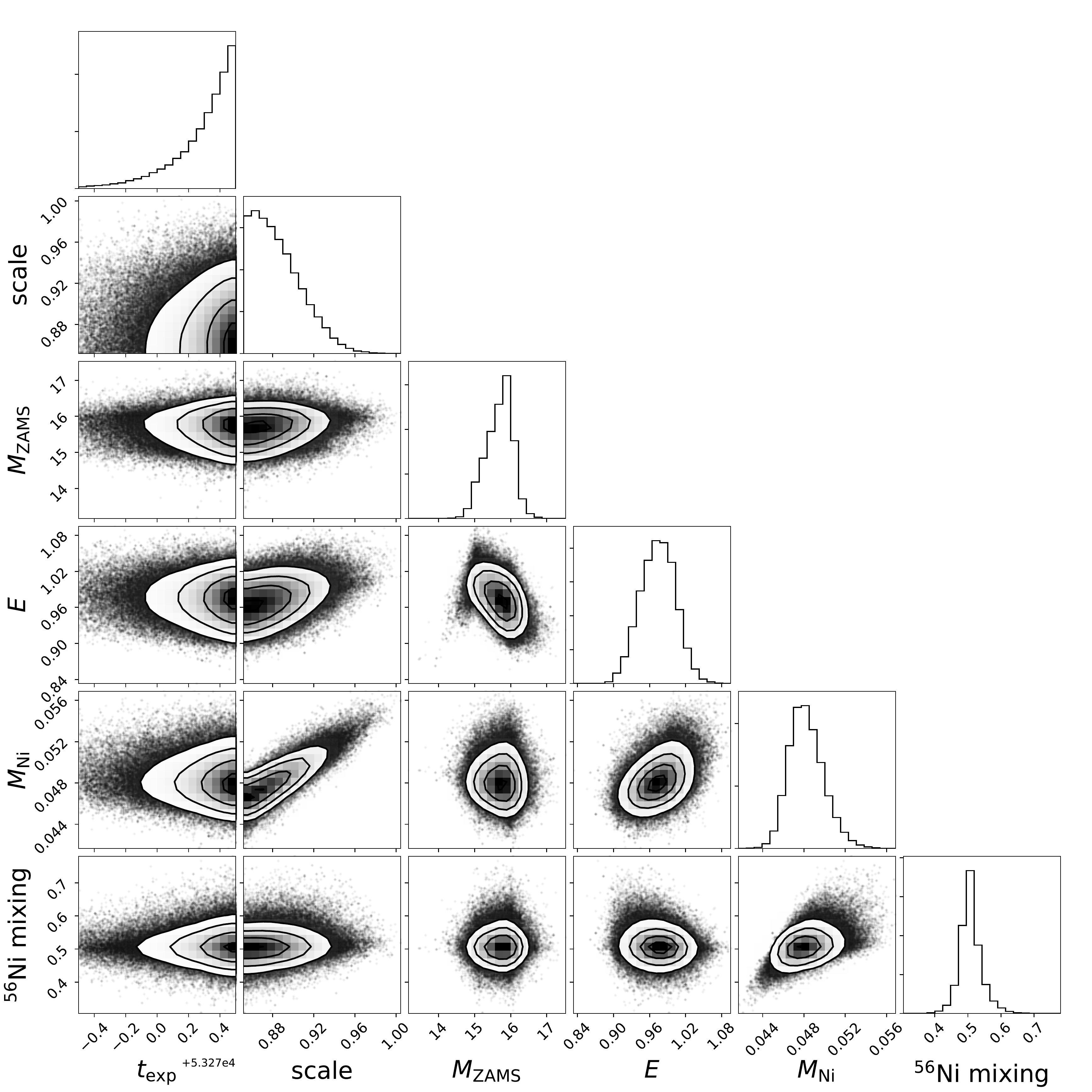}
\caption{Same as for Fig.~\ref{fig:corner_12ec} but for SN~2004et.}
\label{}
\end{figure}

\begin{figure}
\centering
\includegraphics[width=0.5\textwidth]{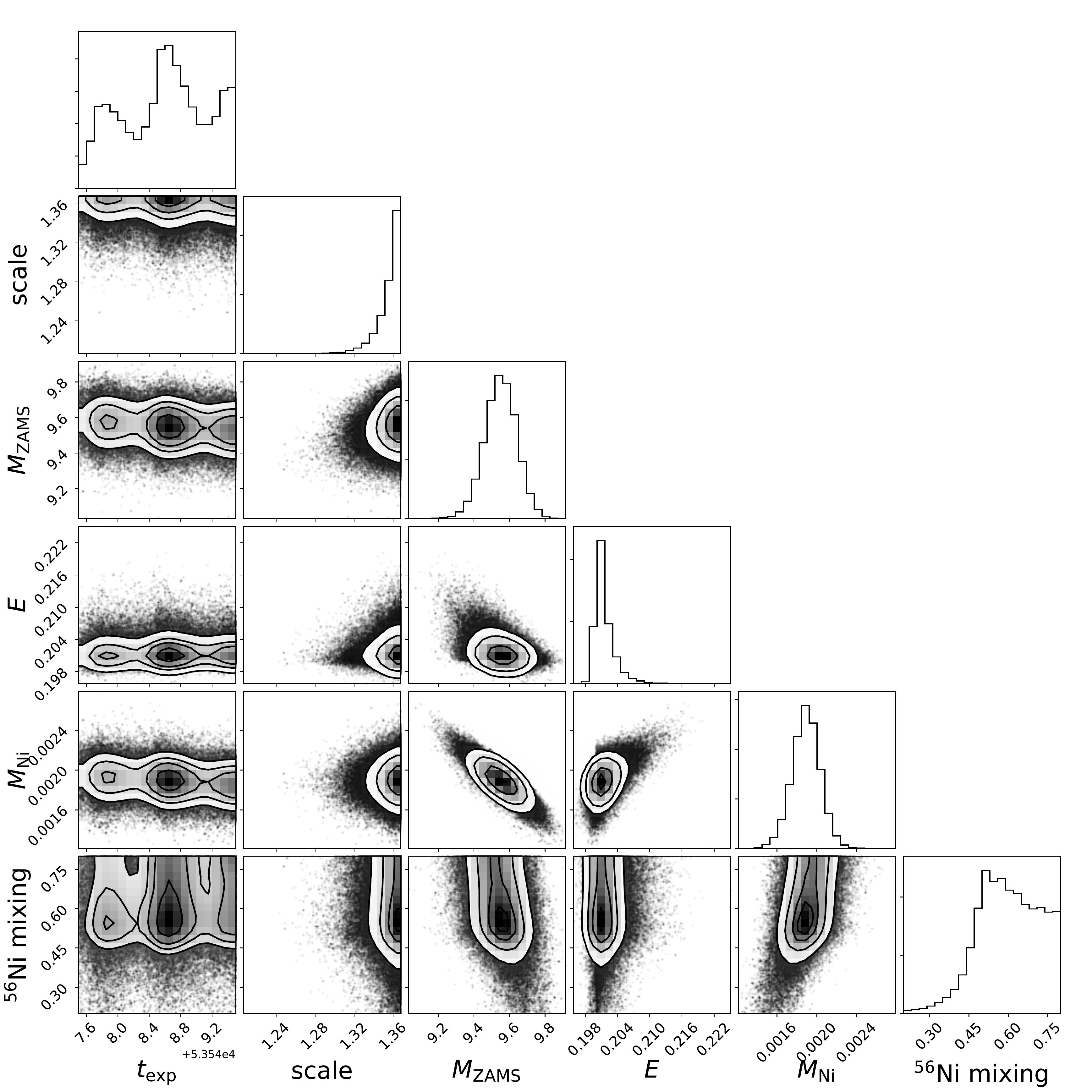}
\caption{Same as for Fig.~\ref{fig:corner_12ec} but for SN~2005cs.}
\label{}
\end{figure}

\begin{figure}
\centering
\includegraphics[width=0.5\textwidth]{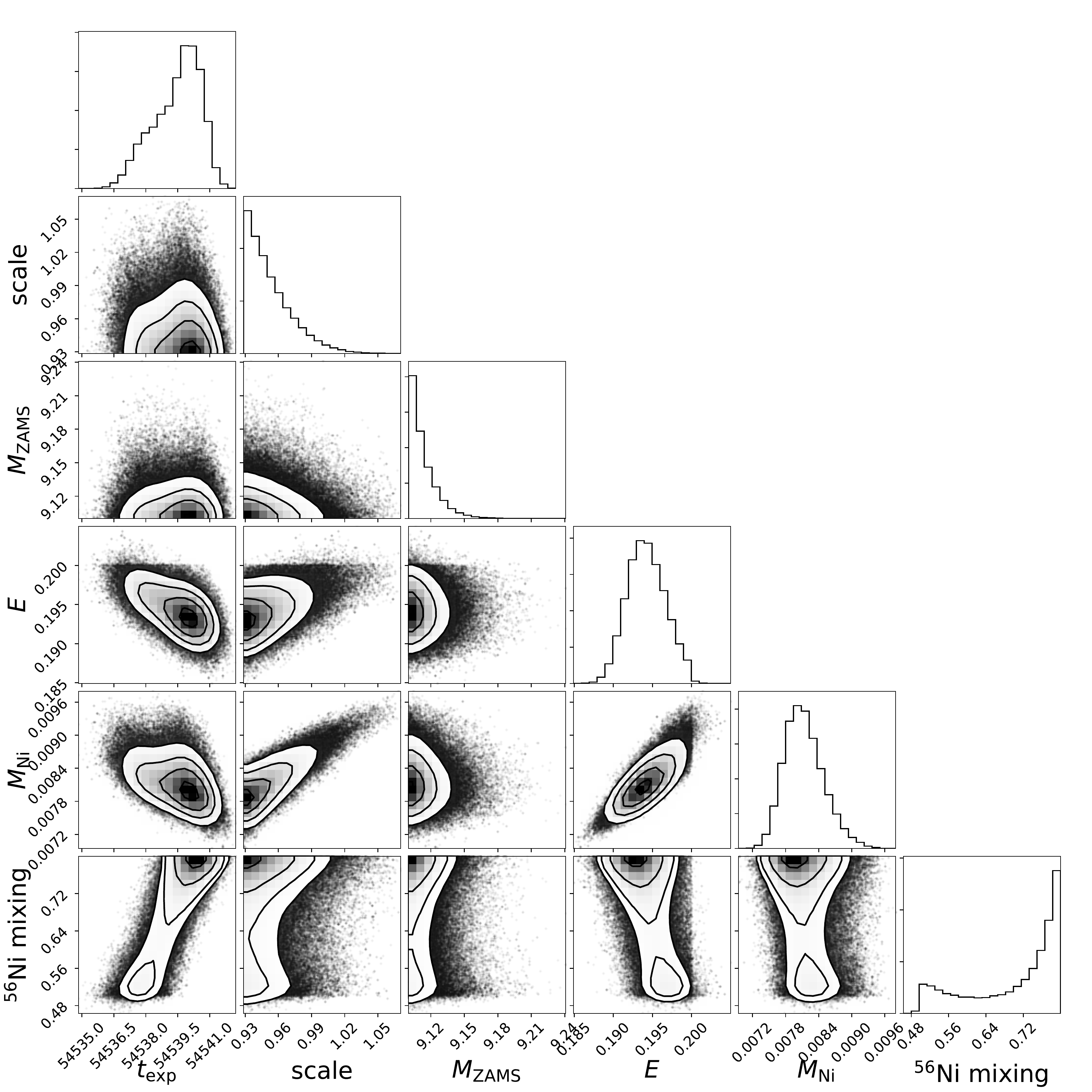}
\caption{Same as for Fig.~\ref{fig:corner_12ec} but for SN~2008bk.}
\label{}
\end{figure}

\begin{figure}
\centering
\includegraphics[width=0.5\textwidth]{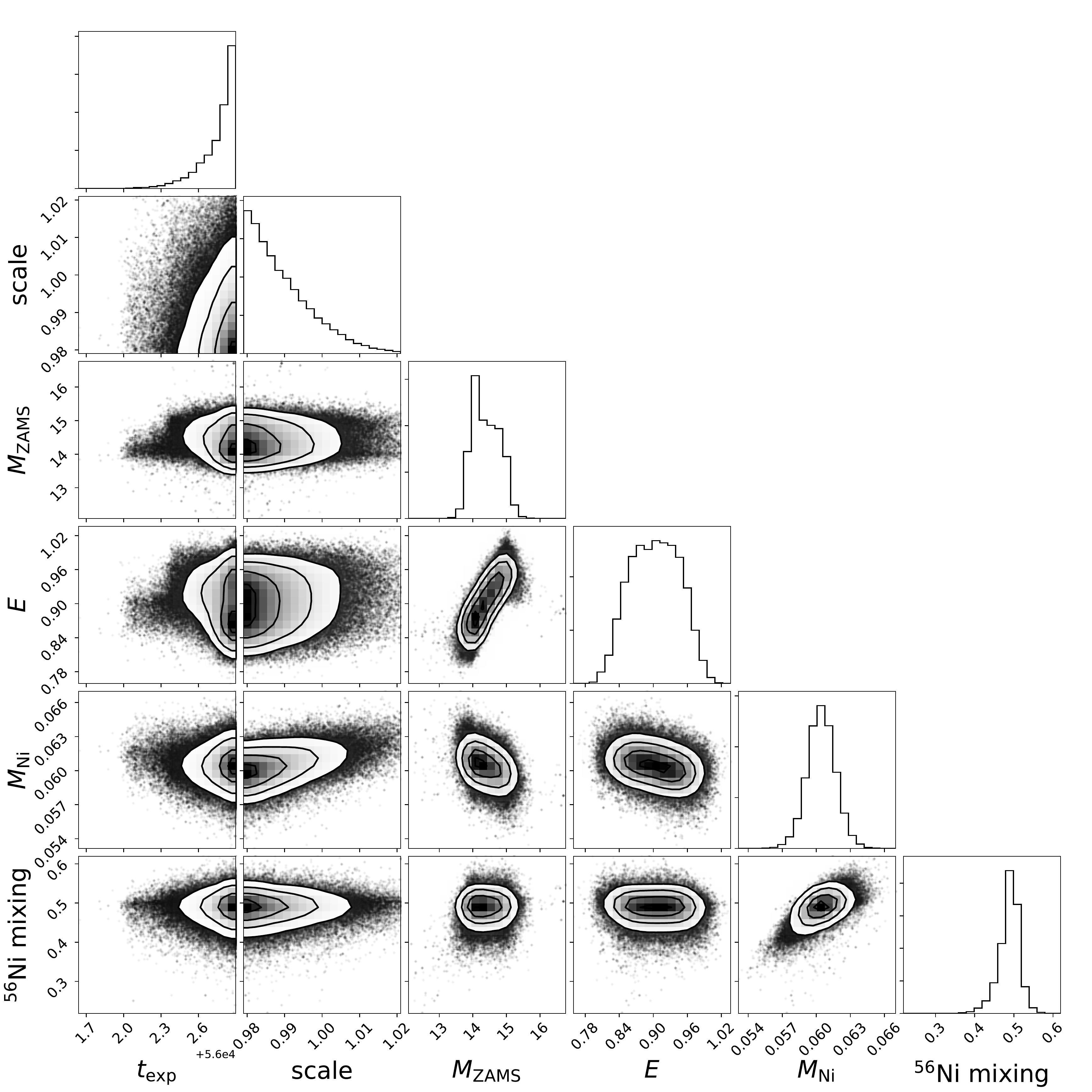}
\caption{Same as for Fig.~\ref{fig:corner_12ec} but for SN~2012aw.}
\label{}
\end{figure}

\begin{figure}
\centering
\includegraphics[width=0.5\textwidth]{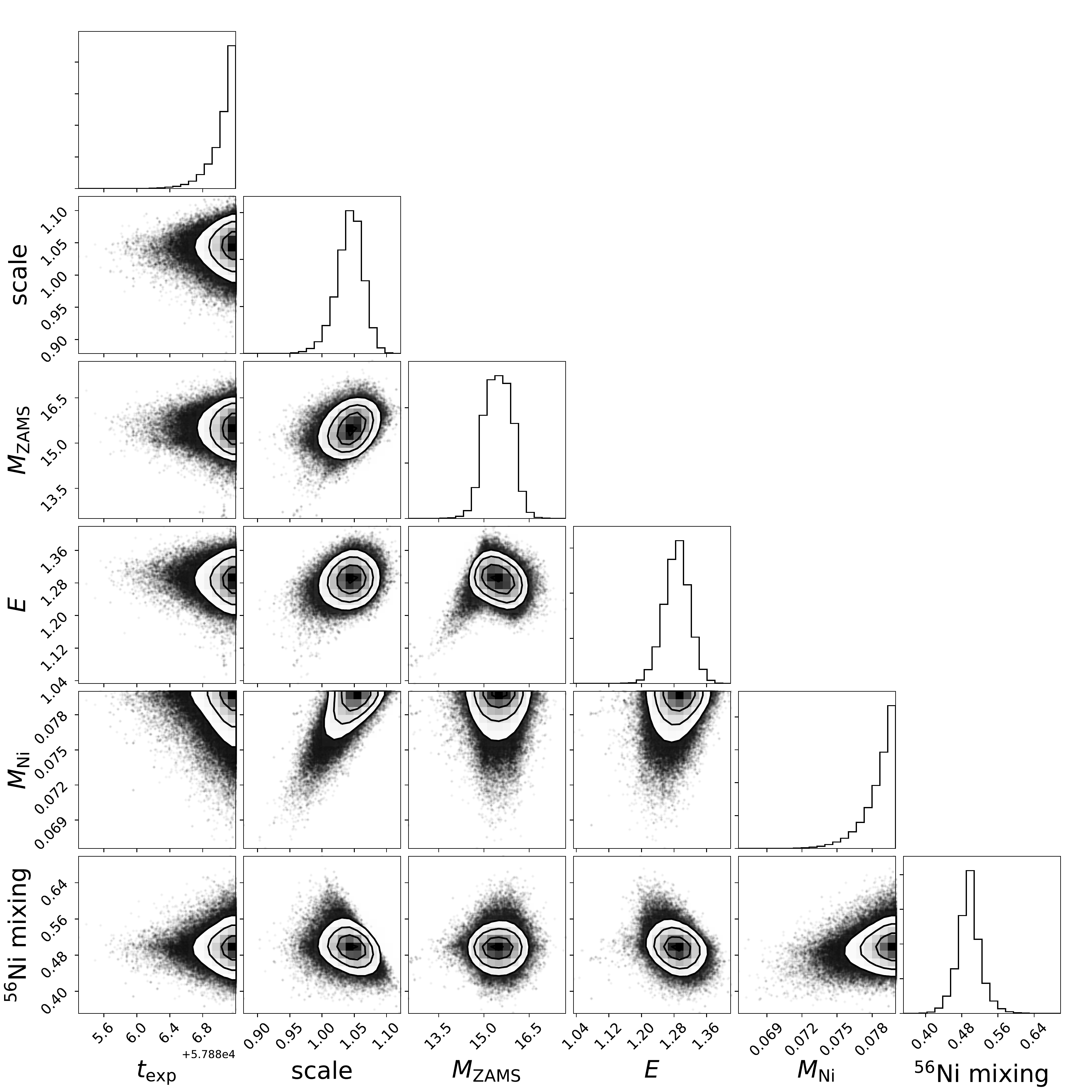}
\caption{Same as for Fig.~\ref{fig:corner_12ec} but for SN~2017eaw.}
\label{}
\end{figure}

\begin{figure}
\centering
\includegraphics[width=0.5\textwidth]{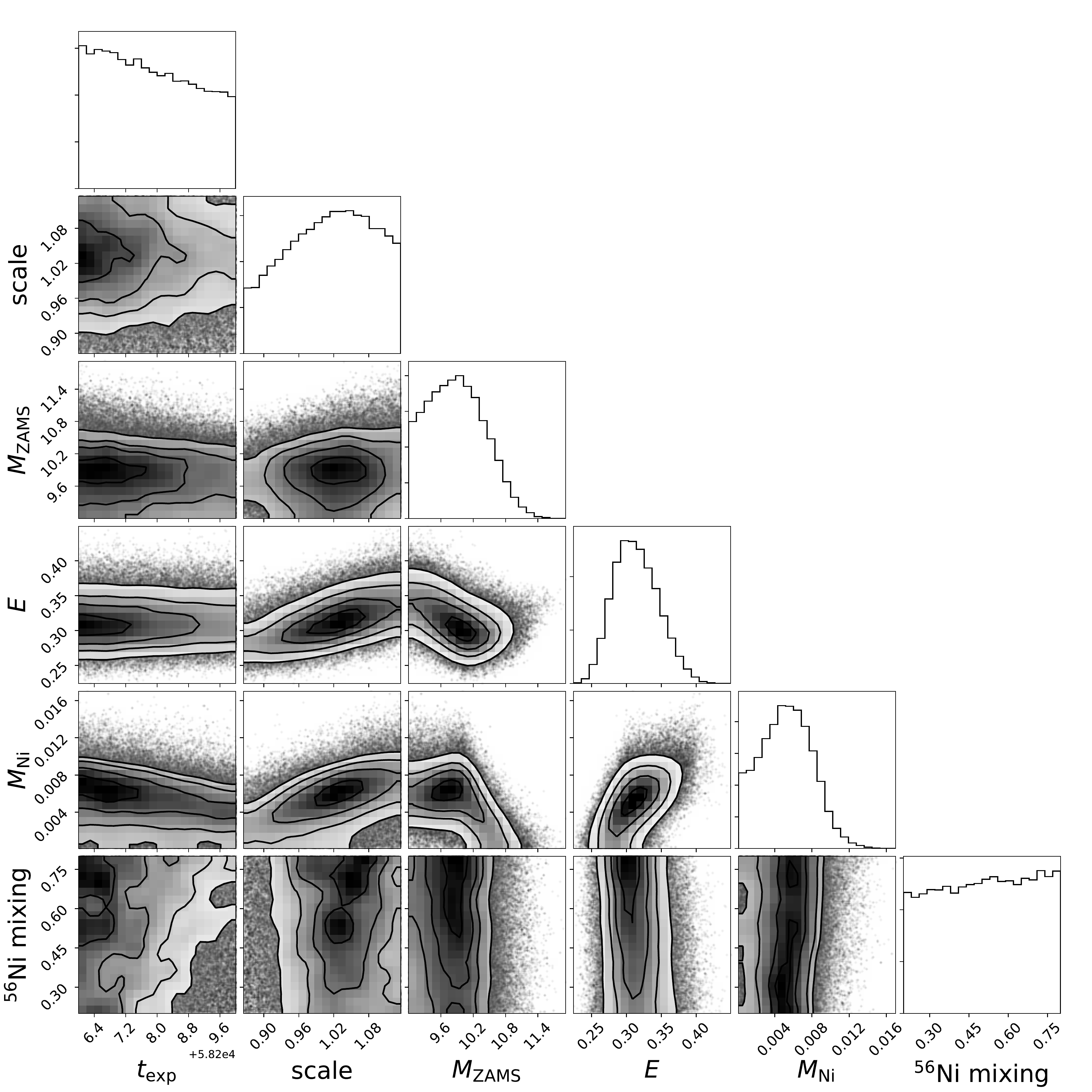}
\caption{Same as for Fig.~\ref{fig:corner_12ec} but for SN~2018aoq.}
\label{}
\end{figure}

\begin{figure}
\centering
\includegraphics[width=0.5\textwidth]{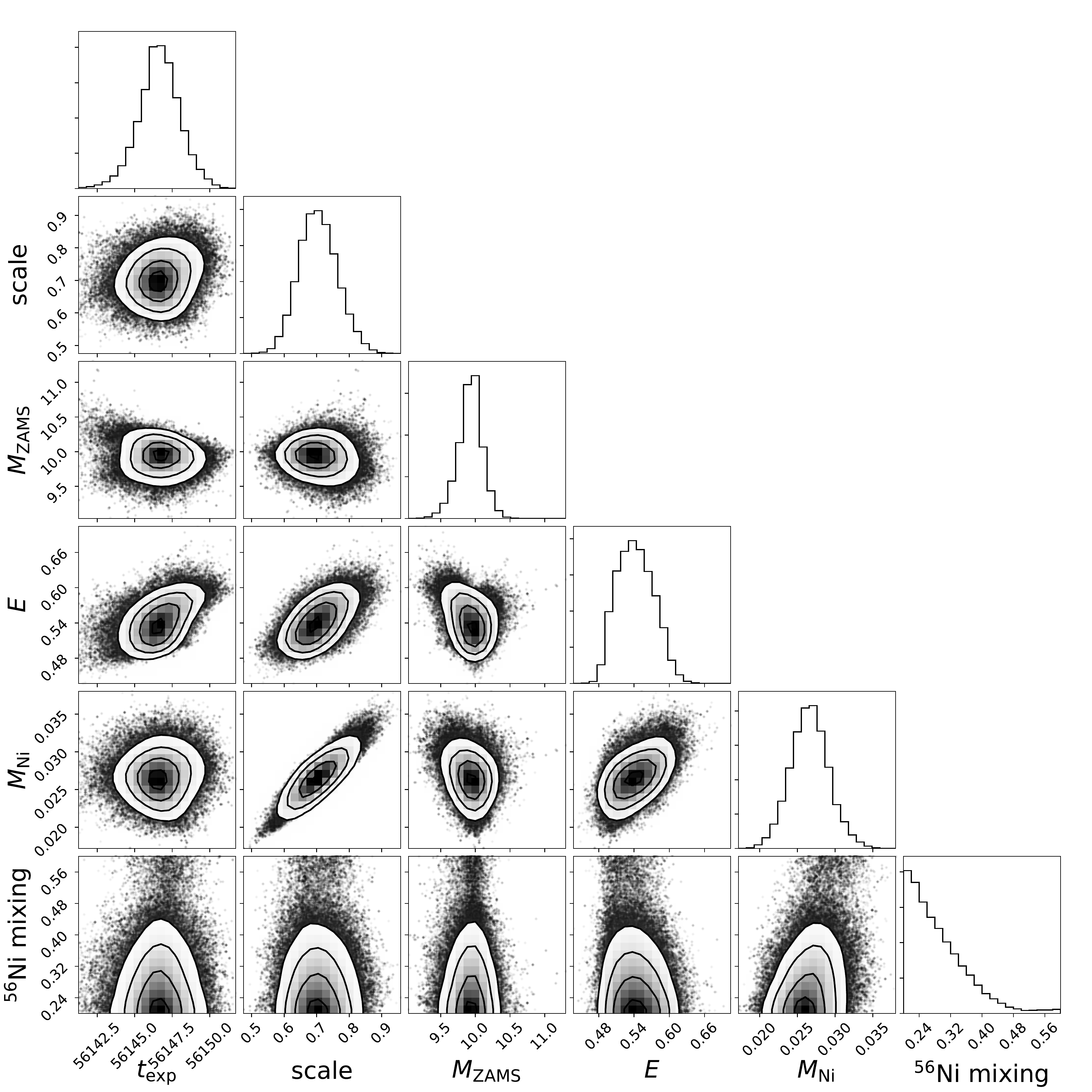}
\caption{Corner plot of the joint posterior probability distribution of the parameters for SN~2012ec when the priors for \te\ and the $scale$ are relaxed.}
\label{fig:corner_12ec_texp_scale}
\end{figure}

\subsection{Autocorrelation and trace plots}

\begin{figure*}
\centering
\includegraphics[width=0.88\textwidth]{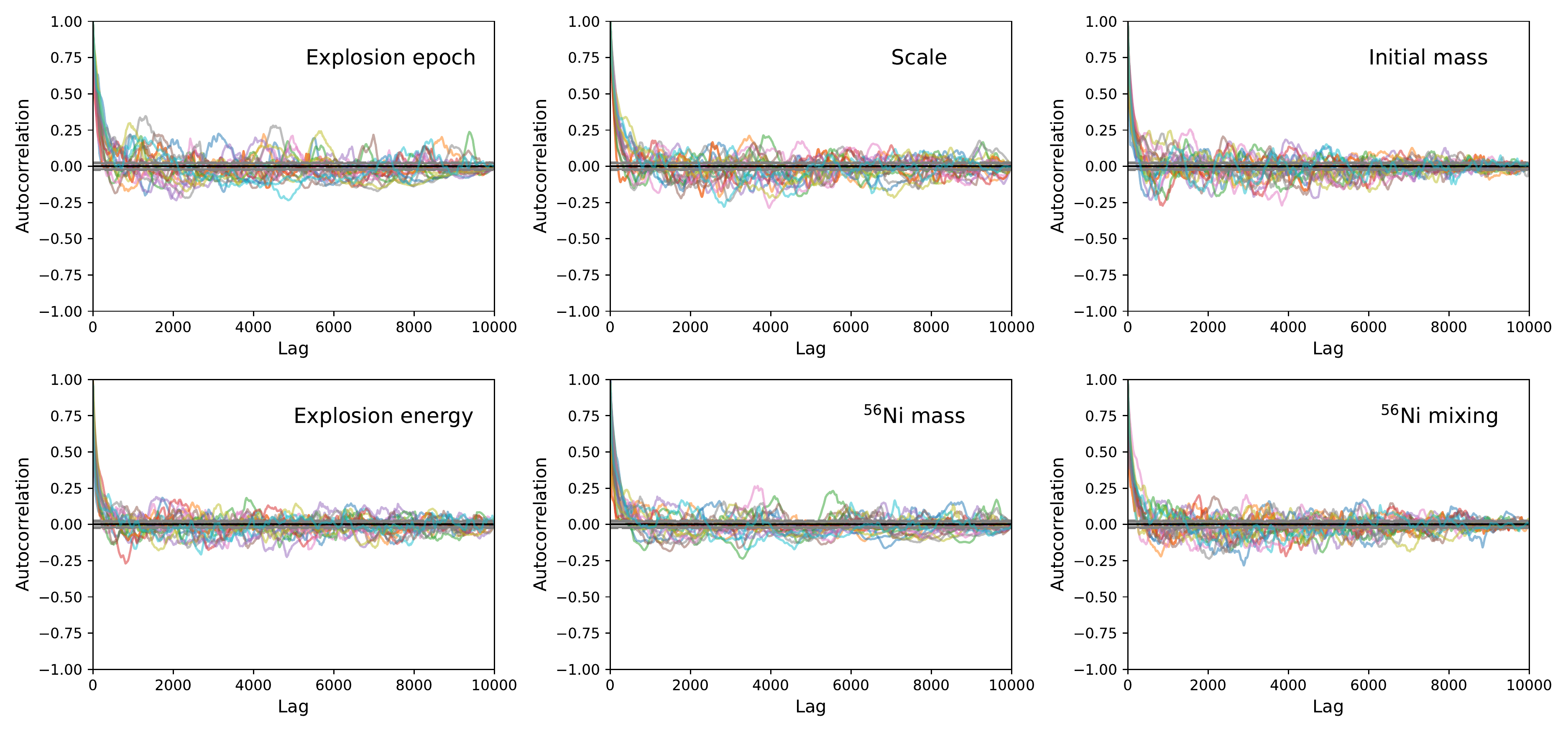}
\caption{Autocorrelation plots for twenty chains randomly chosen using SN~2017eaw as example. Each panel shows the autocorrelation for a different parameter.}
\label{fig:autocorrelation}
\end{figure*}

\begin{figure*}
\centering
\includegraphics[width=0.88\textwidth]{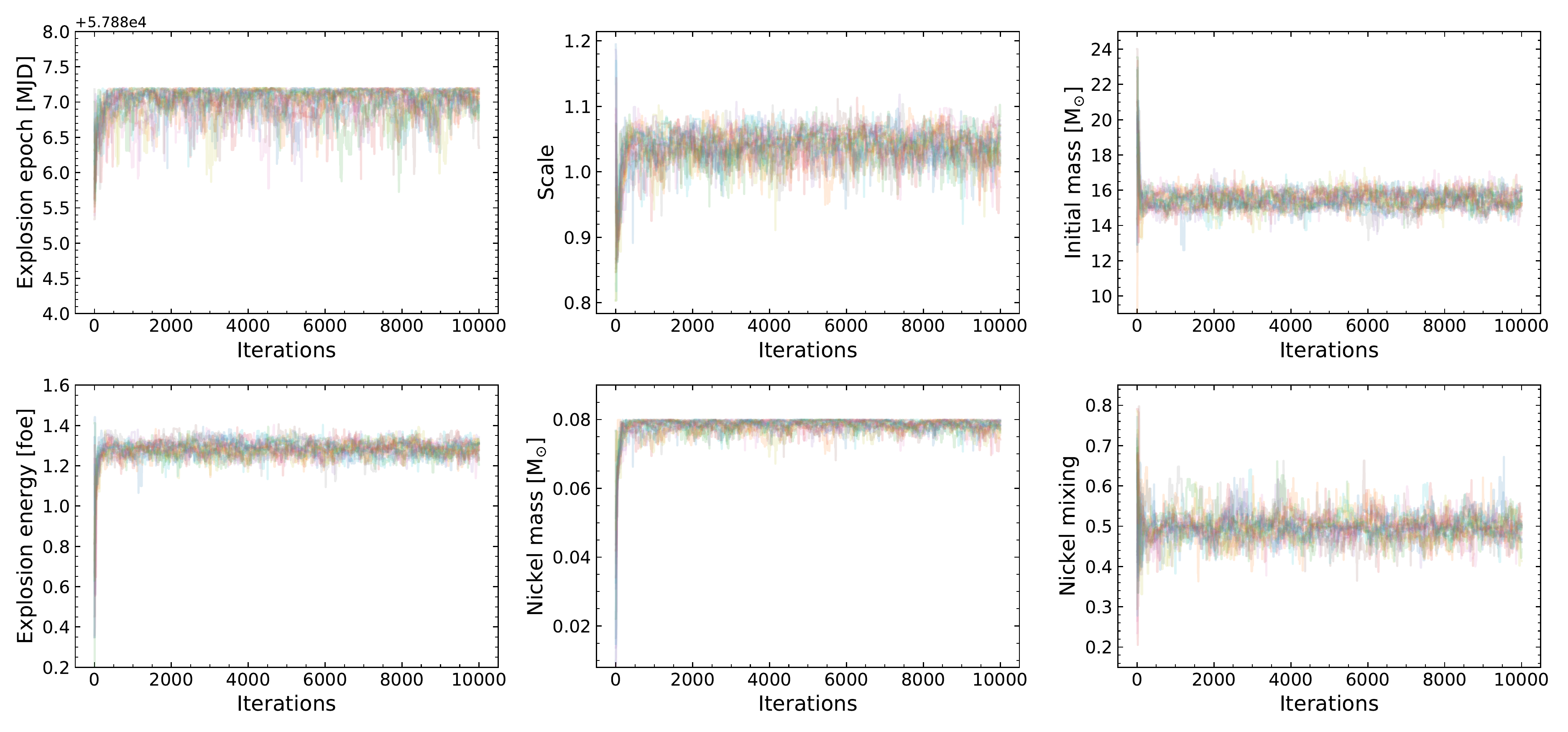}
\caption{Trace plots of the parameters for twenty chains randomly chosen using SN~2017eaw as example. These plots show the evolution of the chains over time.}
\label{fig:traceplot}
\end{figure*}

Here, examples of autocorrelation plots are shown in Fig.~\ref{fig:autocorrelation}. These were performed using the \texttt{autocorrelation\_plot} tool implemented in the \texttt{Python} library \texttt{pandas} \citep{pandas}. Additionally, Fig~\ref{fig:traceplot} shows the trace plots of the MCMC samples. In both cases, we use SN~2017eaw as example.

\subsection{SN~2004et}

\begin{figure*}
\centering
\includegraphics[width=0.85\textwidth]{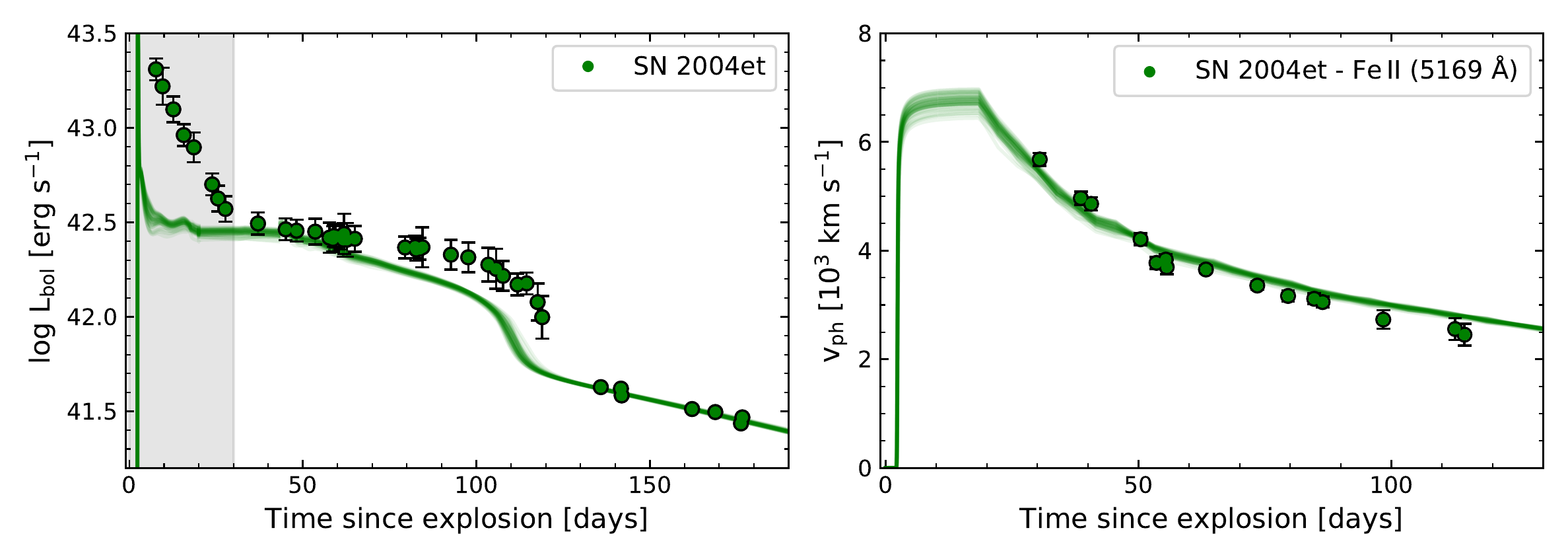}
\caption{Comparison between models and observations for SN~2004et assuming $d$~=~7.73~$\pm$~0.78~Mpc. We show fifty models randomly chosen from the posterior probability distribution. The lack of agreement during the photospheric phase is easily seen. \textit{Left:} bolometric LC. \textit{Right:} evolution of the photospheric velocity. The grey shaded region shows the early data we removed from the fitting. }
\label{fig:sn2004et_badfit}
\end{figure*}

Figure~\ref{fig:sn2004et_badfit} shows models drawn from the posterior distribution of the parameters for SN~2004et when using a distance to the host galaxy of 7.73~$\pm$~0.78~Mpc. The lack of agreement between models and observations is easily seen. The LC models present large discrepancies during the photospheric phase. The models are on average $\sim$0.27~dex fainter and evolve more rapidly.

\end{appendix}

\end{document}